\documentclass[9pt]{revtex4}
\usepackage{amsmath}
\usepackage{amsfonts}
\usepackage{amssymb}
\usepackage{mathrsfs}
\usepackage{graphicx}
\usepackage{subcaption}
\usepackage{multirow}
\usepackage{url}
\usepackage{color}
\newcommand\M{\mathbf{M}}
\newcommand\Q{\mathbf{Q}}
\newcommand\n{\mathbf{n}}
\newcommand\Ivec{\mathbf{I}}
\newcommand\tr{\mathrm{tr}}
\newcommand\atan{\mathrm{atan}}

\begin{document}
\title{Tailored Nematic and Magnetization Profiles on 2D Polygons}
\author{Yucen Han$^{1}$}
\author{Joseph Harris$^{1}$}
\author{Joshua Walton$^{2}$}
\author{Apala Majumdar$^{1}$}
\affiliation{$^1$ Department of Mathematics and Statistics, University of Strathclyde, G1 1XQ, United Kindom.\\
$^2$School of Mathematics and Statistics, University of Glasgow, G12 8QQ, United Kingdom.}

\begin{abstract}
We study dilute suspensions of magnetic nanoparticles in a nematic host, on two-dimensional (2D) polygons. These systems are described by a nematic order parameter and a spontaneous magnetization, in the absence of any external fields. We study the stable states in terms of stable critical points of an appropriately defined free energy, with a nemato-magnetic coupling energy. We numerically study the interplay between the shape of the regular polygon, the size of the polygon and the strength of the nemato-magnetic coupling for the multistability of this prototype system. Our notable results include (i) the co-existence of stable states with domain walls and stable interior and boundary defects, (ii)  the suppression of multistability for positive nemato-magnetic coupling, and (iii) the enhancement of multistability for negative nemato-magnetic coupling.
\end{abstract}

\maketitle
\section{\label{sec:level1}Introduction}
Nematic liquid crystals (NLCs) are classical examples of soft materials that exhibit fluidity and long-range orientational order \cite{de1993physics}. The NLC molecules, typically rod-like in shape, tend to align along locally preferred directions, referred to as  ``directors’’ in the literature, and exhibit orientational order about these directors \cite{de1993physics, stewart2019static, virga1995variational}. Hence, NLCs are intrinsically directional in nature with  direction-dependent optical, mechanical, rheological and electromagnetic responses. The directional dependent properties of NLCs make them the preferred working material of choice for a plethora of electro-optic devices \cite{lagerwall2012new}. In recent years, there has been substantial interest in controlling nematic directors and defects (regions of reduced orientational order where the nematic directors cannot be defined) by embedded inclusions e.g., dispersed colloidal nanoparticles, geometric frustration leading to complex self-assembled structures, new bio-materials,  and topological materials etc. \cite{muvsevivc2006two, bisoyi2011liquid, mur2017magnetic, ackerman2017static}. In this paper, we focus on dilute suspensions of magnetic nanoparticles (MNPs) in a nematic host. Here, the NLC-MNP interactions can lead to a spontaneous magnetization in addition to the nematic directors, in the absence of any external fields. The two effects: magnetization and nematic directors, are coupled by means of a nemato-magnetic mechanical coupling. This coupling is dictated by the the surface treatment of the MNPs \cite{zadorozhnii2008frederiks, potisk2017dynamic}. Some of these composite systems are referred to as ``ferronematics’’  with non-zero net magnetization in the absence of an external magnetic field. Ferronematics were theoretically predicted by the pioneering work of Brochard and de Gennes \cite{brochard1970} with further notable theoretical developments by Burylov and Raikher \cite{burylov1995ferronematics}, among others. Although ferronematic systems were experimentally realized rather early on by Rault, Cladis and Burger \cite{rault1970ferronematics}, stable MNP suspensions have only been recently achieved (see \cite{mertelj2013ferromagnetism, mertelj2017ferromagnetic}).

NLCs have historically relied on their dielectric responses to electric fields for applications because the NLC dielectric anisotropy is several orders of magnitude (e.g., 7 orders of magnitude) larger than the magnetic anisotropy \cite{stewart2019static}. This implies that (unrealistically) large magnetic fields are needed to elicit macroscopic NLC responses to magnetic fields, making it difficult to exploit the magneto-mechanical and magneto-optic properties of NLCs.  The addition of MNPs to a NLC host can substantially increase the magnetic susceptibility of the suspension \cite{zadorozhnii2008frederiks}, and influence phase transition temperatures and other material properties, all of which are largely determined by the surface anchoring on the MNP surfaces. The anchoring depends on the NLC properties \cite{kopvcansky2005anchoring}, or the particle coating of MNPs \cite{chen1983observation}. In confined geometries, the nematic director and the spontaneous magnetization can be tailored through geometric frustration and boundary effects (see e.g., \cite{bisht20}). This yields novel possibilities for NLC devices operated by magnetic fields, biaxial ferronematics and chiral ferronematics that could be used for optics, telecommunications, microfluidics, smart fluids, and diagnostics to name a few \cite{mertelj2017ferromagnetic, sahoo2015magnetodielectric, potisk2017dynamic, hess2015optical}.

In \cite{BishtKonark2019Mnia} and \cite{bisht20}, the authors study a dilute suspension of MNPs in a one-dimensional NLC-filled channel and a NLC-filled 2D square, respectively. They report exotic stable morphologies for the nematic director and the associated magnetization profile, without any external fields. They report the co-existence of stable states with interior nematic defects, interior magnetic vortices, magnetic domain walls that separate \emph{ordered polydomains} i.e., two distinct domains with different magnetizations, and states with defects pinned to the square vertices. These results demonstrate the immense potential of 2D polygons for tailored multistability in ferronematic-type systems, which would be inaccessible in generic confined NLC systems (see e.g., \cite{RobinsonMartin2017Fmtc}, \cite{luo2012multistability}). We build on this work by studying dilute suspensions of MNPs in a nematic host, on 2D regular polygons without external magnetic fields, as a natural generalisation of the work on square wells in \cite{bisht20}. A dilute suspension refers to a uniform suspension of small MNPs (on the nanometer scale with length greater than the diameter) such that the average distance between a pair of distinct MNPs is much greater than the MNP size, and the total volume fraction of suspended MNPs is small. In this limit, it can be mathematically proven, using homogenization techniques, that the MNP-interactions are ``small" compared to other effects, and the NLC-MNP interactions are captured by an ``effective energy". This effective NLC-MNP energy depends on the shape and size of the MNPs, the surface anchoring energies and the nemato-magnetic coupling \cite{calderer2014effective, canevari2020design}. 2D polygons are an excellent approximation to shallow three-dimensional (3D) wells with a 2D polygon cross-section, such that the well height is much smaller than the polygon edge length. From a modelling perspective, it is reasonable to assume that the structural details are invariant across the well height and it suffices to model the ferronematic profiles on the 2D polygonal cross-section; this reduced 2D approach can be rigorously justified (see \cite{GolovatyDmitry2017DRft}, \cite{wang2019order}). Boundary conditions are a crucial consideration for confined systems. We impose fixed/Dirichlet tangent boundary conditions for the nematic director on the polygon edges, and these boundary conditions create a natural mismtach for the nematic director at the polygon vertices, making them natural candidates for defect sites \cite{walton2018nematic, luo2012multistability, han2020reduced}. Tangent boundary conditions are well accepted for confined NLC systems both experimentally and theoretically; see \cite{tsakonas2007multistable}. We impose a fixed topologically non-trivial tangent boundary condition for the spontaneous magnetization of the suspended MNPs; this is a purely theoretical choice for the time being. For a dilute system, it is reasonable to assume that the boundary conditions for the magnetization follow the tangent boundary conditions for the nematic director. This choice of the boundary condition naturally leads to interior magnetic vortices, offering a wonderful playground for exploring exotic solution landscapes of these ferronematic systems.
From an experimental perspective, in \cite{ShuaiM2016Slca}, the authors argue that tangent boundary conditions for the spontaneous magnetization can arise from energetic considerations. We speculate that the boundary conditions for the magnetization could be controlled by applying an external magnetic field to fix the orientation and position of the MNPs on the boundaries, followed by the removal of the magnetic field, although this is largely open to the best of our knowledge. There are multiple choices of boundary conditions for the nematic director and the magnetization, (including free boundary conditions for the magnetization, or weak anchoring effects) but our choice of Dirichlet tangent boundary conditions offers rich possibilities, that could guide future experimental studies on these lines.

There are two macroscopic order parameters: (i) the nematic order parameter: the Landau-de Gennes (LdG) $\Q$-tensor order parameter which encodes both the nematic director, $\n$, and the degree of nematic ordering about $\n$; and (ii) a polar order parameter, described by the averaged spatial magnetization vector, $\M$, of the suspended MNPs without external magnetic fields. We do not account for the volume fraction of the MNPs as in \cite{burylov1995ferronematics}, since we work with dilute uniform suspensions that have a small volume fraction; also see the phenomenological approaches in \cite{potisk2017dynamic, potisk2018magneto}. We model the experimentally observable profiles as minimizers of an appropriately defined energy, as in \cite{bisht20}, \cite{BishtKonark2019Mnia}, which in turn builds on the free energy descriptions in \cite{mertelj2017ferromagnetic}, \cite{potisk2017dynamic}, \cite{potisk2018magneto}. The proposed free energy has three essential contributions: a conventional nematic free energy; 
a magnetic energy that coerces a preferred value of $|\M|$ as in \cite{potisk2017dynamic, potisk2018magneto, burylov2013magnetically} and includes a Dirichlet energy density term to penalise arbitrary rotations between $\M$ and $-\M$; and crucially a nemato-magnetic coupling energy parameterised by a coupling parameter $c$. The Dirichlet energy density for $\M$ is, mathematically speaking, a regularisation term and does not introduce new physics into the problem for judicious parameter choices. In the dilute limit, the nemato-magnetic coupling energy is the \emph{homogenized limit} of a Rapini-Papoular type of surface anchoring energy on the MNP surfaces \cite{canevari2020design}. In the dilute limit, we do not see the individual MNPs but rather account for the collective NLC-MNP interactions, mediated by the surface anchoring energies, in terms of this effective nemato-magnetic coupling energy. In principle, one could use homogenization methods to compute effective nemato-magnetic coupling energies for arbitrary MNP shapes, and other types of MNP surface anchoring energies e.g.,  we expect the coupling energies to be different for platelet-shaped MNPs but we adopt the simplest approach here. For $c>0$, the nematic director $\n$ and $\M$ prefer to be either parallel or anti-parallel to each other and the Dirichlet energy density for $\M$ regularises the $\M$ profile. For $c<0$, $\n$ and $\M$ tend to be perpendicular to each other in the polygon interior and this naturally creates fascinating boundary layers near the polygon edges. From \cite{calderer2014effective}, both cases of positive and negative $c$ are physically relevant, and $c$ may be an experimentally tunable parameter. 

There are five key dimensionless parameters in the model: $N$ – the number of sides of the confining geometry; $\ell_1$ and $\ell_2$ – rescaled elastic constants associated with the Dirichlet energy density of $\Q$ and $\M$, respectively which are inversely proportional to the polygon edge length; $\xi$ – which is a magnetic coherence length that weighs the relative importance of the nematic and magnetic energies; and $c$ is the nemato-magnetic coupling parameter which determines the co-alignment of $\n$ and $\M$. The coupling parameter $c$ can be related to the volume fraction, size and shape of the MNPs, and the strength of the MNP-NLC interactions.  
We consider $N=4, 5, 6$ in this manuscript i.e., a square, a regular pentagon and hexagon, respectively.
Specifically, we numerically compute the energy-minimizing $(\Q,\M)$ profiles for different values of $N$, $c$ and $\ell$ , along with bifurcation diagrams for positive and negative values of $c$ that track the energy-minimizing stable and non energy-minimizing (unstable) solution branches. 
Mathematically, this corresponds to solving a system of four coupled nonlinear partial differential equations, subject to Dirichlet conditions for $\Q$ and $\M$. The NLC system (with $c=0$) has been well described in \cite{han2020reduced} on 2D polygons, where the authors demonstrate a unique $Ring$ solution profile with a unique nematic point defect at the center, which is the generic stable solution for  polygons except for the square for large $\ell$. The authors find at least $\left[\frac{N}{2} \right]$ stable states for small $\ell$. A key question is - how does this picture respond to the NLC-MNP coupling, captured by the parameter $c$? There are various new solutions for these nemato-magnetic systems, as will be described in the sections below, which are not reported for the $c=0$ case in \cite{han2020reduced}. Some notable findings for positive $c$ concern the coexistence of stable $(\Q, \M)$ profiles with nematic defects pinned at the polygon vertices and magnetic domain walls along polygonal diagonals and polygon edges, that separate distinct domains of magnetization; along with stable \emph{$Peppa$}- $(\Q, \M)$ profiles. The $Peppa$-branches have two $+1/2$ interior nematic defects and a magnetic vortex at the center. We note that magnetic domain walls are difficult to find with either increasing $N$ or increasing $c \in \left(0, \infty \right)$. The picture with negative $c$ is more complex -  we effectively double the number of stable states for small $\ell$, compared to the results in \cite{han2020reduced} for $c=0$. These stable states are distinguished by vertex defects for $\Q$ and vertex vortices for $\M$, so that the multistability is strongly enhanced with increasing $N$, for $c<0$. Additionally, we find stable $(\Q, \M)$-profiles, labelled as \emph{$Peppa_{in}$, $Peppa_{out}$} solutions, with complex permutations of interior $+1/2$-nematic defects and magnetic vortices, for small $\ell$. We compute bifurcation diagrams for representative values $c=0.25$ and $c=-0.25$, as a function of $\ell$ and $N$, to capture the solution branches as a function of the polygon edge length encoded in $\ell$, and to illustrate defect-induced multistability.

The paper is organized as follows. In Section II, we outline the theoretical framework and the governing equilibrium equations for this ferronematic-type system. In Section III, we present a comprehensive numerical study of the equilibria in a square complemented by some analysis in two asymptotic limits. In Sections IV and V, we present  numerical results for a hexagon and a pentagon, respectively and we summarise the principal conclusions and directions for further research in Section VI.

\section{\label{sec:model}Model Formulation}
\maketitle
We study partially ordered 2D systems on a square, pentagon and hexagon, with nematic orientational order and polar magnetic order, motivated by recent studies of dilute ferronematic suspensions \cite{bisht20}.  More specifically, the domain $\Omega$ is a re-scaled regular $N$-polygon centered at the origin; we note that the physical edge length $L$ has been absorbed into the phenomenological parameters as will be described below (see \cite{bisht20}). The polygon vertices are defined  by
\begin{gather}
v_k=  \left(\cos\left(\frac{2\pi(k-1)}{N}\right),\sin\left(\frac{2\pi(k-1)}{N}\right)\right)
\end{gather}
for $k=1,\dots,N$.
The polygon edges are labelled counterclockwise as $C_1,\dots,C_N$, such that $C_1$ connects $v_1$ to $v_2$, and so on.

These 2D systems have two order parameters - a rescaled LdG $\Q$-tensor order parameter and a 2D magnetization vector,  $\M=(M_1,M_2)$ which is the polar order parameter.  In 2D, the reduced LdG $\Q$-tensor order parameter can be written as \cite{han2020reduced}
\begin{gather}
\Q=S(2\n\otimes\n -	\Ivec),
\end{gather}
where the nematic director, $\n=(\cos\theta,\sin\theta)^T$ describes the preferred in-plane alignment of the nematic molecules, and $S$ is the scalar order parameter which measures the degree of orientational order about the planar director. For a rigorous justification of the reduced 2D LdG approach, see \cite{GolovatyDmitry2017DRft}. Therefore, $\Q$ has two independent components: 
\begin{gather}
\Q=
\begin{pmatrix}
Q_{11} & Q_{12} \\
Q_{12} & - Q_{11}\\	
\end{pmatrix},
\end{gather}
where $Q_{11}=S\cos2\theta$ and $Q_{12}=S\sin2\theta$.
In this framework, we will not detect biaxial regions since $\tr{\Q}^3=0$ and we have $\tr{\Q}^2=|\Q|^2=2Q_{11}^2+2Q_{12}^2$. 
We assume that $\M$ is the spontaneous magnetization induced by the MNPs with an internal magnetic moment, which interacts with $\n$ through surface anchoring conditions on the MNP surfaces. $\M$ has variable magnitude: magnetic vortices are defined by $|\M| =0$, and defective regions are identified by reduced values of $|\M|$. As described in the Introduction, we assume a dilute suspension of MNPs in a nematic host, 
and the total re-scaled and dimensionless free energy is given by
\begin{align}
\mathcal{F}[\Q,\M]=&\int_\Omega\frac{1}{4}\left\{\ell_1|\nabla\Q|^2+\frac{1}{4}|\Q|^4-|\Q|^2\right\}\,\mathrm{dA}\nonumber\\&+\int_\Omega\frac{\xi}{2}\left\{\ell_2|\nabla\M|^2+\frac{1}{2}|\M|^4-|\M|^2\right\}\,\mathrm{dA}\nonumber\\
&-\int_\Omega\frac{c}{2}\M^T\Q\M\,\mathrm{dA}, \label{energy}
\end{align}
where the first line is the nematic energy, the second line is the magnetic energy, the last line is the effective nemato-coupling energy. 
We work with low temperatures so that the bulk favours an ordered nematic and magnetic phase with $|\Q|\neq 0, |\M| \neq 0$. The total bulk potential is
\begin{gather}
    \frac{1}{4}|\Q|^4-|\Q|^2 + \xi|\M|^4 -2\xi |\M|^2 - 2c\M^T\Q\M;
\end{gather}
the corresponding stationary points (in terms of $c$ and $\xi$) are computed in \cite{dalby21}.
 
There are $4$ parameters above as stated in the Introduction: $\ell_1$,  $\ell_2$, the magnetic coherence length $\xi$, and the nemato-magnetic coupling parameter, $c$. $\ell_1$ is defined to be the ratio of a material-dependent length scale and the physical edge length  $L$ i.e., $\ell_1 =\frac{K}{|A| L^2}$ where $K$ is the nematic elastic constant, $|A|$ is proportional to the absolute temperature and $L$ is the edge length.
Further, from the form of the nemato-magnetic coupling energy density, $-2 c \M^T \Q \M$, positive $c$ favours $\left(\n\cdot \M\right)^2 = 1$ and negative $c$ favours $\n \cdot \M = 0$ (see \cite{bisht20}). The magnetic Dirichlet energy density is a regularisation energy that smoothens the $\M$ profiles and prevents degeneracy of energy minimizers. 
As is standard in the calculus of variations, the physically observable equilibria are local or global minimizers of (\ref{energy}), subject to the boundary conditions. However, unstable critical points of (\ref{energy}) play a crucial role in transition pathways between distinct equilibria, see \cite{kusumaatmaja2015free}. The  critical points (stable or unstable) of (\ref{energy}) are solutions of the associated Euler-Lagrange equations:
\begin{align}
\ell_1\Delta Q_{11}&=\tilde{Q}Q_{11}-\frac{c}{2}(M_1^2-M_2^2),\label{EL1}\\
\ell_1\Delta Q_{12}&=\tilde{Q}Q_{12}-cM_1M_2,\label{EL2}\\
\xi \ell_2\Delta  M_1&=\xi\tilde{M}M_1-c(Q_{11}M_1+Q_{12}M_2),\label{EL3}\\
\xi \ell_2\Delta   M_2&=\xi\tilde{M}M_2-c(Q_{12}M_1-Q_{11}M_2),\label{EL4}
\end{align}
where $\Delta$ is the two-dimensional Laplacian operator, and $\tilde{Q}=\left(\frac{1}{2}\tr{\Q}^2-1\right)$ and $\tilde{M}=(|\M|^2-1)$. 
The phenomenological parameters, $\ell_1, \ell_2, \xi$ and $c$ are typically estimated from experimentally measured quantities but the available data is limited, in the presence of external magnetic fields \cite{potisk2017dynamic}. We investigate the sensitivity of the solution landscapes with respect to the re-scaled elastic constants and $c$. The elastic constants depend on the temperature, material-dependent constants and the physical length $L$ of the domain, and hence, they are tunable parameters. The parameter $c$ depends on the ratios of material-dependent constants and the strength of the NLC-MNP interactions, so could also be a tunable parameter. The last parameter, $\xi$ is the ratio of NLC material constants and MNP-dependent constants, and again could be reasonably tuned in moderate regimes. For simplicity, we fix $\xi=1$, assume that the re-scaled nematic and magnetic elastic constants satisfy $\ell_1 = \ell_2= \ell$, unless stated otherwise. These choices improve the efficiency of our numerical procedure and allow us to capture the complex solution landscapes. For a dilute system, we expect $\ell_2$ to be (much) smaller than $\ell_1$ but the qualitative properties of the bifurcation diagrams remain unchanged compared to the $\ell_1=\ell_2$ case, with shifted bifurcation points. 

As stated in the Introduction, we assume fixed Dirichet tangent boundary conditions for $\Q$ and $\M$, which requires both the nematic director, $\n$, and $\M$ to be tangent to the edges of $\Omega$. 
We assume that $\M$ rotates by $2\pi$ radians around $\partial \Omega$, which is a topologically non-trivial boundary condition that naturally induces an interior magnetic vortex. 
Regarding $\n$, we assume $\n$ is tangent to the edges $C_k$, and there is a natural mismatch at the vertices, $v_k$. More specifically, the square domain has vertices at $\left(-0.5, \pm 0.5 \right)$ and $\left(+0.5, \pm 0.5 \right)$ such that
\begin{eqnarray}
    \label{eq:square_bcs}
   && Q_{11b} = 1 ~\textrm{on $y=\pm 0.5$}; Q_{11b} = -1 ~\textrm{on $x=\pm 0.5$} \nonumber \\
   && Q_{12b} = 0 ~\textrm{on $x=\pm 0.5$, $y=\pm 0.5$;} \nonumber \\
   && (M_{1b},M_{2b}) = (-1, 0)~\textrm{on $y=-0.5$}; \nonumber\\
   && (M_{1b},M_{2b}) = (1, 0)~\textrm{on $y=0.5$}; \nonumber \\
   && (M_{1b},M_{2b}) = (0, 1)~ \textrm{on $x = -0.5$}; \nonumber\\ 
   && (M_{1b},M_{2b}) = (0, -1) ~\textrm{on $x = 0.5$.}\nonumber\\
\end{eqnarray}
For a pentagon and a hexagon with $N=5$ or $N=6$, we specify the boundary conditions on the edges $C_k$ for $k=1 \dots N$, as follows:
\begin{align}
(Q_{11b},Q_{12b})=\left(-\cos\left(\frac{2\pi(2k-1)}{N}\right),\sin\left(\frac{2\pi(1-2k)}{N}\right)\right),\label{BC1}
\end{align}
and 
\begin{gather}
(M_{1b},M_{2b})=\left(\sin\left(\frac{\pi(2k-1)}{N}\right),-\cos\left(\frac{\pi(2k-1)}{N}\right)\right).	\label{BC2}
\end{gather}
 We numerically compute the solutions of the system (\ref{EL1})--(\ref{EL4}), subject to the Dirichlet boundary conditions (\ref{eq:square_bcs}) ((\ref{BC1})--(\ref{BC2}) in the pentagon/hexagon), which are necessarily critical points of (\ref{energy}). We use the DOLFIN library \cite{LoggWells2010a} from the popular open-source computing platform FEniCS \cite{AlnaesBlechta2015a} which allows us to solve the weak form of the Euler-Lagrange equations, in a suitable finite element function space. This solver uses Newton's method to find weak solution of the Euler-Lagrange equations \cite{LoggMardalEtAl2012a}, and is unlikely to converge to an unstable solution. Due to the high multiplicity of the solutions, convergence may be highly sensitive to the choice of initial condition. In the following figures, we plot $|\Q|$, labelled by the colour chart, and the nematic director, $\n$, by white lines where $\n$ is given by 
\begin{gather}\label{eq:n}
\n=(\cos\theta,\sin\theta),\qquad \theta=\frac{1}{2}\atan2(Q_{12},Q_{11});	
\end{gather}
and $|\M|$ labelled by the colour chart, and the white arrows describe the magnetic orientation $(M_1,M_2)/ |\M|$ for $|\M|\neq 0$. We study the stability of the solutions by numerically calculating the smallest real eigenvalue $\lambda_1$ of the Hessian of the energy (\ref{energy}) with four degrees of freedom $Q_{11}$, $Q_{12}$, $M_1$, and $M_2$ using the LOBPCG ((locally optimal block preconditioned conjugate gradient) method \cite{KnyazevAndrewV2001TtOP}). If $\lambda_1$ is positive, the solution is stable.
The case $c=0$, has been studied in \cite{han2020reduced}, and the authors report the WORS (Well Order Reconstruction Solution) on a square, and the $Ring$ branch on other regular polygons, for large $\ell$. These solutions bifurcate to $Diagonal$ ($D$), $Rotated$ ($R$) solutions on a square; $Para$, $Meta$ and $Ortho$ solutions branches on regular polygons with $N>4$, as $\ell$ decreases. The numerical computation of bifurcation diagrams requires continuation techniques, for which we first locate different stable solutions.
We find a new solution, $Peppa$, with stable interior $+1/2$-nematic defects, for $c = 0.25$, by taking the $D$ solution (for $c = 0$) as the initial condition for the Newton’s method. The new solutions $Peppa_{in}$ ($Peppa_{out}$) for $c = -0.25$ are obtained by taking the profiles $\left(\mathbf{Q}, (M_2, -M_1)\right)$ ($\Q, (-M_2, M_1)$) as initial conditions where ($\mathbf{Q}, \mathbf{M} = (M_1,M_2))$ is the $Peppa$ solution for $c = 0.25$. Once the $Peppa$, $Peppa_{in}$, $Peppa_{out}$ solutions are computed for $c = \pm 0.25$, we perform a decreasing (increasing) $\ell$ sweep 
for $c=\pm 0.25$, to compute the corresponding bifurcation diagrams.

\section{Solution Landscape on a Square}

We first recall the essential results for a square domain, for $c=0$ from \cite{RobinsonMartin2017Fmtc}, where the authors track the solutions of (\ref{EL1})-(\ref{EL2}) subject to (\ref{eq:square_bcs}), as a function of the square edge-length, $L$, at a fixed temperature. The re-scaled elastic constant, $\ell \propto \frac{1}{L^2}$, at fixed temperature.
For large $\ell$ or small $L$ ($\ell > 0.1$ or $L< 10^{-7}$m approximately), there is a unique WORS \cite{KraljSamo2014Orpi}, distinguished by a pair of mutually orthogonal defect lines along the square diagonals (with $\Q \approx 0$). The WORS is a special case of the more general $Ring$ solution for $N$-polygons reported in \cite{han2020reduced}, and exists for all $\ell >0$ on a square domain with tangent boundary conditions (\ref{eq:square_bcs}). As $\ell$ decreases, the WORS loses stability and bifurcates into two stable diagonal, $D$ solutions, for which $\n$ aligns along one of the square diagonals in the interior. The $D$ solutions have two diagonally opposite \emph{splay} vertices, such that the corresponding $\n$ has a splay pattern near the splay vertex. As $\ell$ further decreases, there is a further bifurcation point with two unstable BD solution branches bifurcating from the WORS branch. The BD solutions have two defect lines parallel to a pair of opposite square edges and the BD solutions further bifurcate into 4 unstable rotated solutions ($R$) as $\ell$ decreases. The nematic director, $\n$, rotates by $\pi$ radians between a pair of opposite square edges for a $R$ solution, and there are $4$ rotationally equivalent $R$ solutions related by a $\frac{\pi}{2}$-rotation. In contrast to the $D$ solutions, each $R$ solution has a pair of \emph{splay} vertices connected by a square edge. The $R$ solutions gain stability as $\ell$ decreases, and for $\ell$ small enough ($\ell < 10^{-3}$ or $L > 10^{-6}$m approximately), there are six distinct stable nematic equilibria; $2$ $D$ solutions and $4$ $R$ solutions. 

\begin{figure}
\centering
    \begin{subfigure}{0.74\columnwidth}
        \centering
        \includegraphics[width=\textwidth]{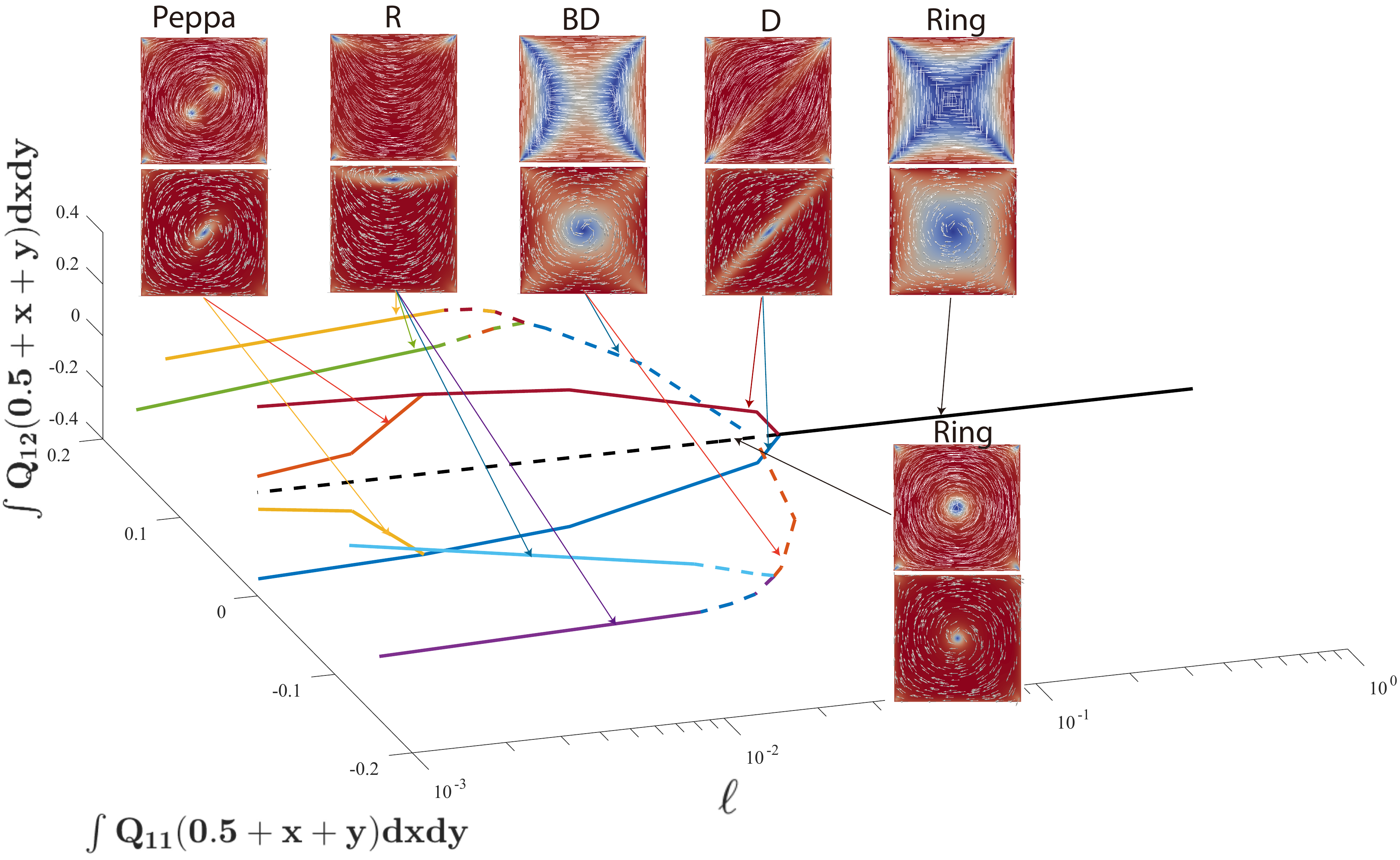}
    \end{subfigure}
    \begin{subfigure}{0.25\columnwidth}
        \centering
        \includegraphics[width=\textwidth]{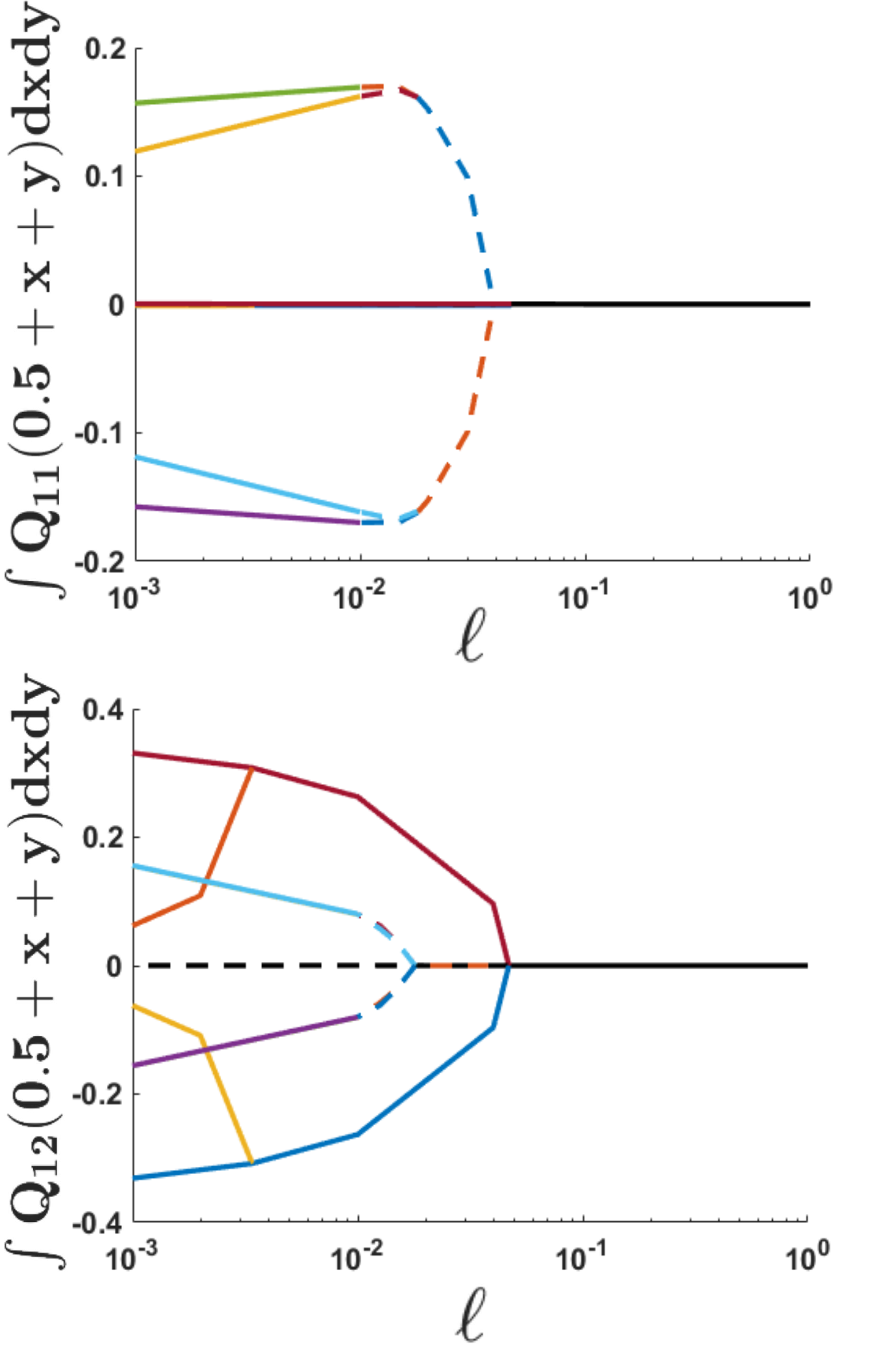}
    \end{subfigure}
        \caption{Bifurcation diagram for (\ref{energy}) on a square domain with $c = 0.25$. Left : plot of $\int Q_{11}\left(0.5+x+y\right)\textrm{d}x\textrm{d}y$, $\int Q_{12}\left(0.5+x+y\right)\textrm{d}x\textrm{d}y$ versus $\ell$; right: orthogonal 2D projections of the full 3D plot.}
        \label{fig:bifurcation_diagram_25}
\end{figure}

\begin{figure}
		\centering
        \includegraphics[width=0.5\columnwidth]{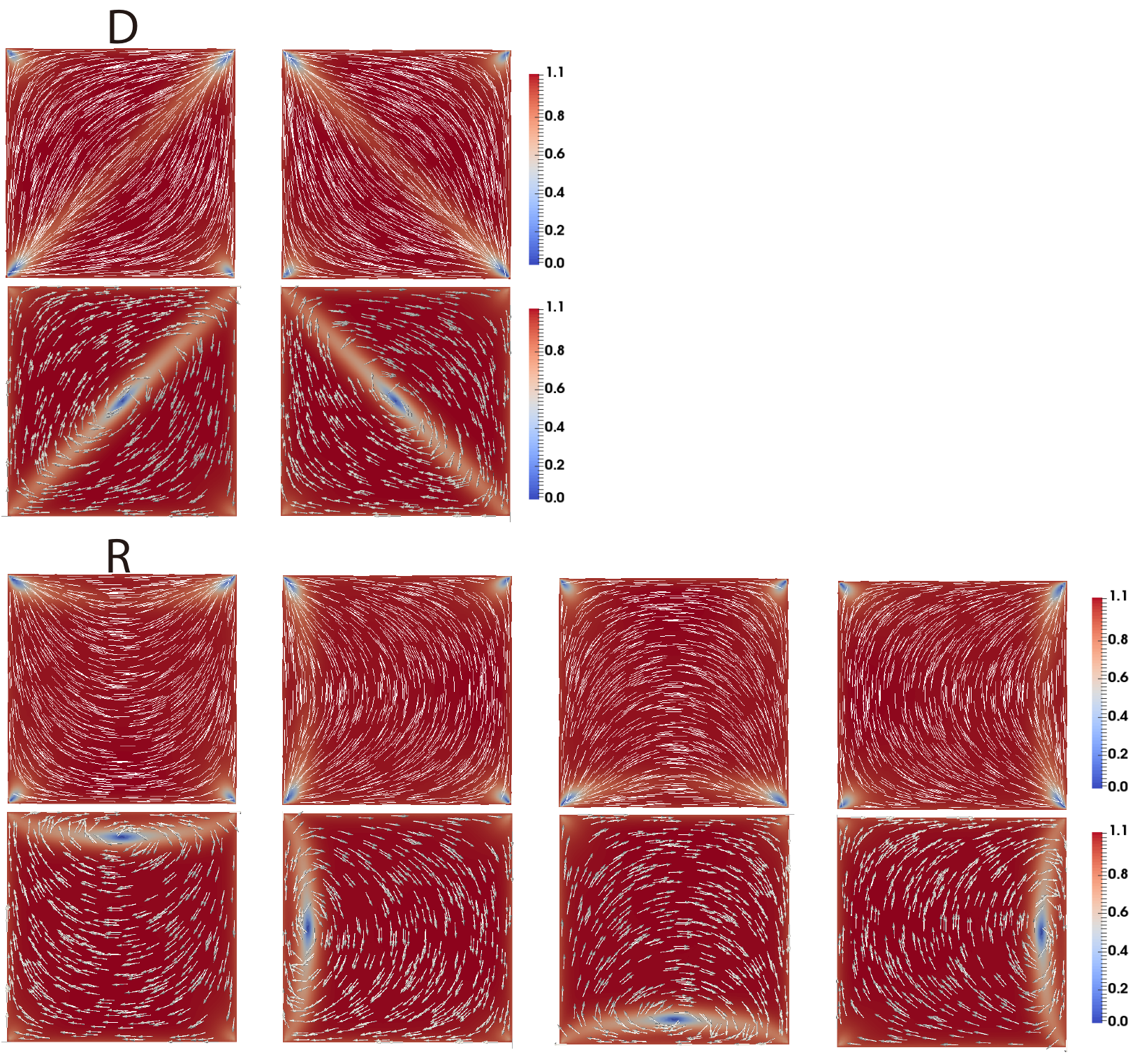}
        \caption{The plots of two $D$ and four $R$ solutions with $\ell=10^{-3}$ and $c=0.25$. Top row: In the nematic profile, the vector $\mathbf{n}$ in \eqref{eq:n} is represented by white lines and the order parameter $|\mathbf{Q}|/\sqrt{2} = \sqrt{Q_{11}^2 + Q_{12}^2}$ is labelled by the colour chart. Bottom row: In the magnetization profile, $|\M|$ is labelled by the colour chart, and the white arrows describe the magnetic orientation $(M_1,M_2)/ |\M|$ for $|\M|\neq 0$. All subsequent figures have nematic profiles in the top row and magnetization profiles in the bottom row, and have the same color map and interpretation of the lines and vectors.}
        \label{fig:25_D_R}
\end{figure}

The qualitative features of the bifurcation diagram are unchanged for $c>0$; see Figure~\ref{fig:bifurcation_diagram_25}. We distinguish between the distinct solution branches by defining two measures, $\int Q_{11}(0.5+x+y) \mathrm{dxdy}$ and $\int Q_{12}(0.5+x+y) \mathrm{dxdy}$, and plot these measures versus $\ell$ for the different solutions. Solid lines represent stable solution branches and dashed correspond to unstable branches. For $c=0.25$ and for $\ell$ large, we have a unique solution of the system (\ref{EL1})-(\ref{EL4}), subject to the boundary conditions (\ref{eq:square_bcs}). The unique $\Q$-solution is the WORS reported in \cite{KraljSamo2014Orpi} and the unique $\M$-solution has a magnetic vortex of degree $+1$ (determined by the degree of the boundary conditions) at the square centre. 
This solution branch exists for all $\ell>0$, but loses stability as $\ell$ decreases. As $\ell$ decreases, the WORS loses the cross structure and collapses into a $Ring$ solution with a circular nematic defect, analogous to the magnetic vortex, at the square centre. We refer to this solution branch, which is unique and globally stable for $\ell$ large enough, as the \emph{Ring branch}. As $\ell$ decreases, the $Ring$ branch loses stability and bifurcates into two stable $D$ solutions (with regards to the $\Q$-solutions). The corresponding $\M$-profiles have domain walls (with reduced $|\M|$) along the corresponding square diagonals. As we will explain below, these domain walls correspond to a $\pi$-rotation in the $\M$-vector. As $\ell$ decreases further, the unstable $Ring$ branch bifurcates into two unstable BD branches (with regards to the $\Q$-solutions). Each BD solution bifurcates into two unstable $R$ solutions, which gain stability when $\ell$ is small enough. The $\M$-solutions, corresponding to the stable $R$ solutions, exhibit a domain wall along the square edge with the two splay vertices. We observe a novel feature for $c>0$ - the stable $D$ solutions bifurcate into two $Peppa$ solution branches that have two +1/2-nematic defects along the square diagonal, for $\ell$ small enough. The $Peppa$ solutions, with pairs of interior nematic defects, are stable for $c=0.25$; we speculate that they exist for all $c>0$ but are unstable for $c=0$. The corresponding $\M$-profiles have a smeared out vortex along the line connecting the nematic defect pair in the $Peppa$ solutions. This is an interesting example of how nemato-magnetic coupling stabilises domain walls in $\M$ (from the $D$ and $R$ solutions) and interior point defects in $\Q$, in terms of the $Peppa$ solutions. Plots of the 2 $D$ solutions and 4 $R$ solutions for $c=0.25$, 
are presented in Figure \ref{fig:25_D_R}.  

\begin{figure}
\centering
    \begin{subfigure}{0.74\columnwidth}
        \centering
        \includegraphics[width=\textwidth]{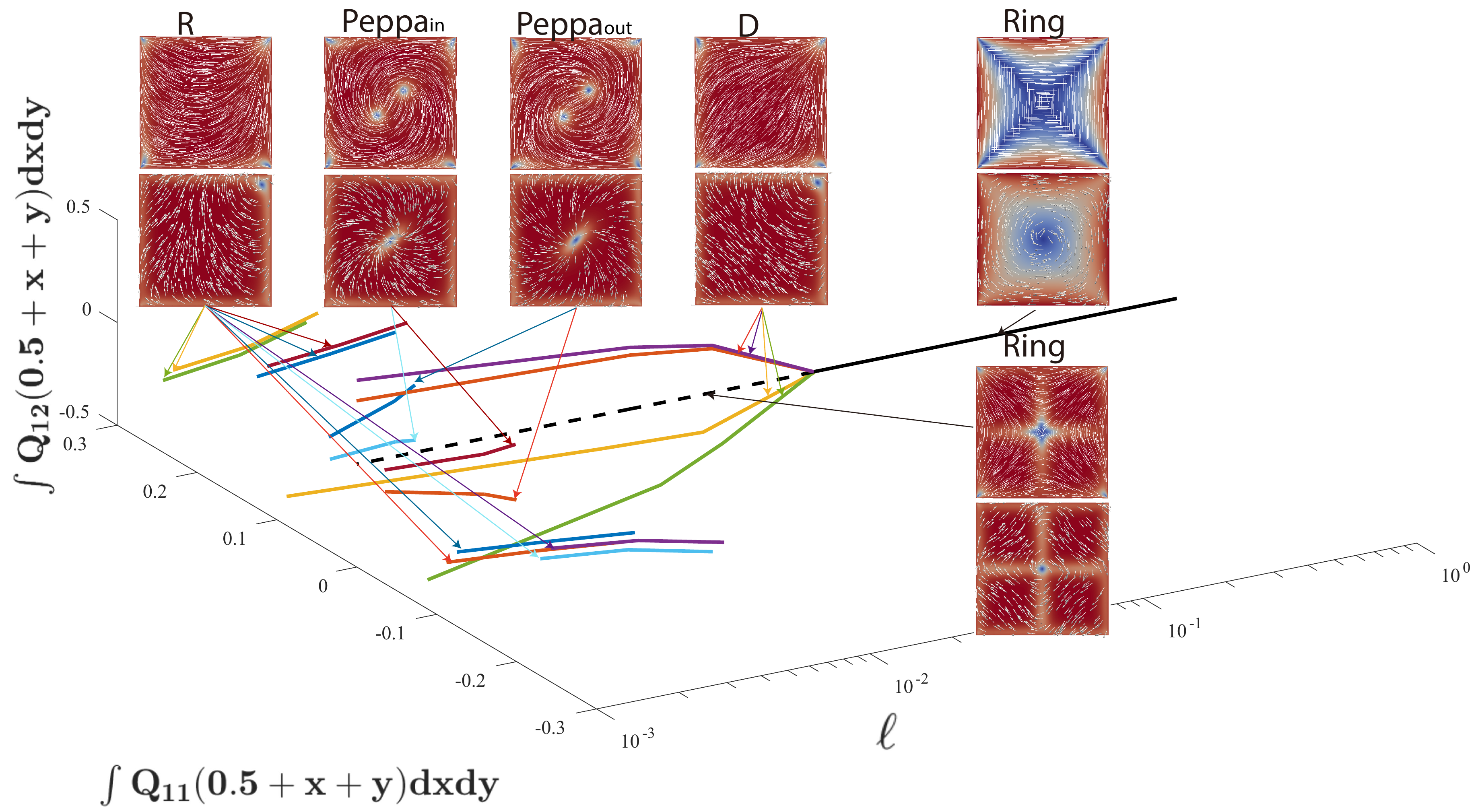}
    \end{subfigure}
    \begin{subfigure}{0.25\columnwidth}
        \centering
        \includegraphics[width=\textwidth]{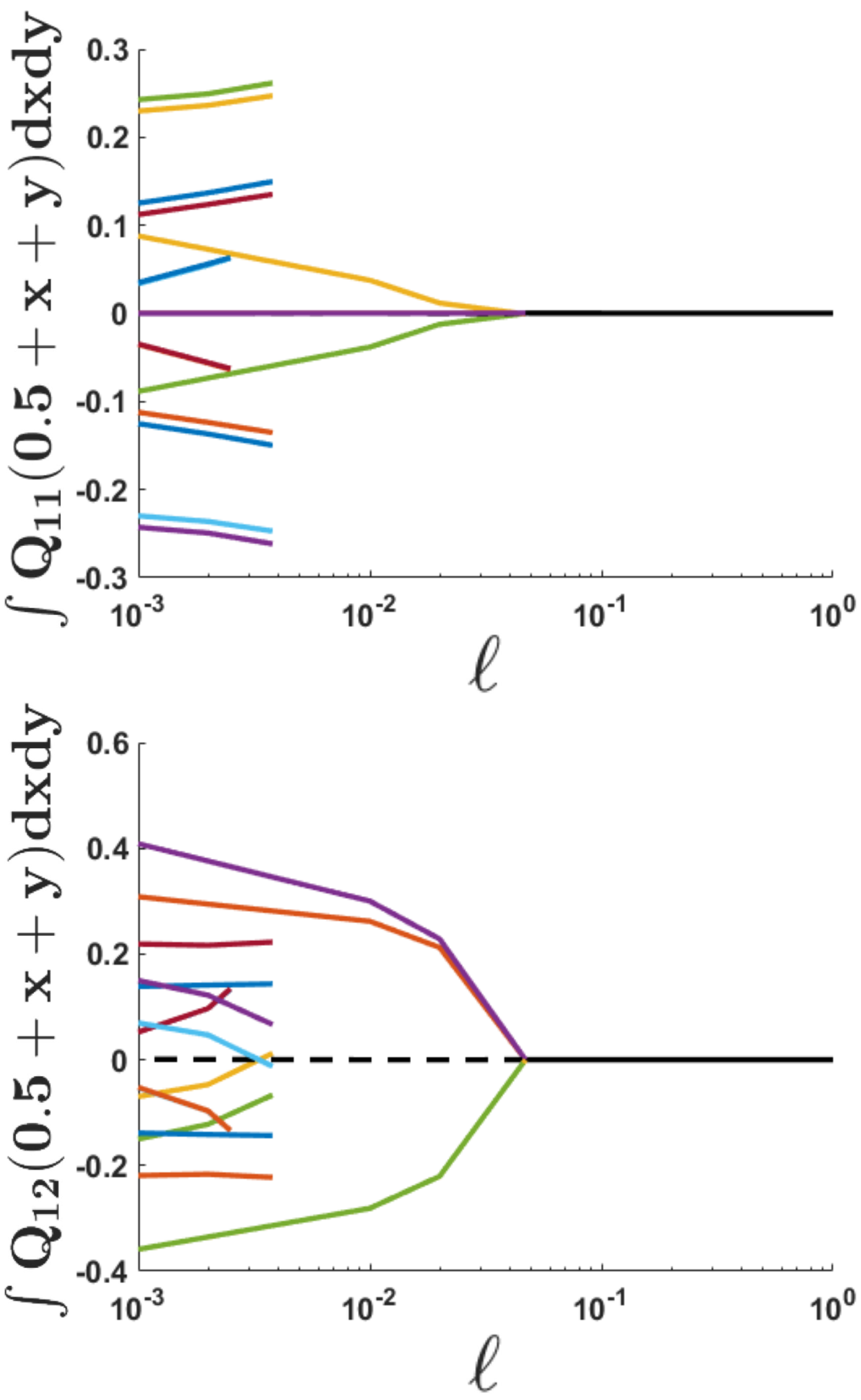}
    \end{subfigure}
        \caption{Bifurcation diagram for (\ref{energy}) on a square domain with $c = -0.25$. Left : plot of $\int Q_{11}\left(0.5+x+y\right)\textrm{d}x\textrm{d}y$, $\int Q_{12}\left(0.5+x+y\right)\textrm{d}x\textrm{d}y$ versus $\ell$; right: orthogonal 2D projections of the full 3D plot.}
        \label{fig:bifurcation_diagram_neg25}
\end{figure}

\begin{figure}
\centering
        \includegraphics[width=\columnwidth]{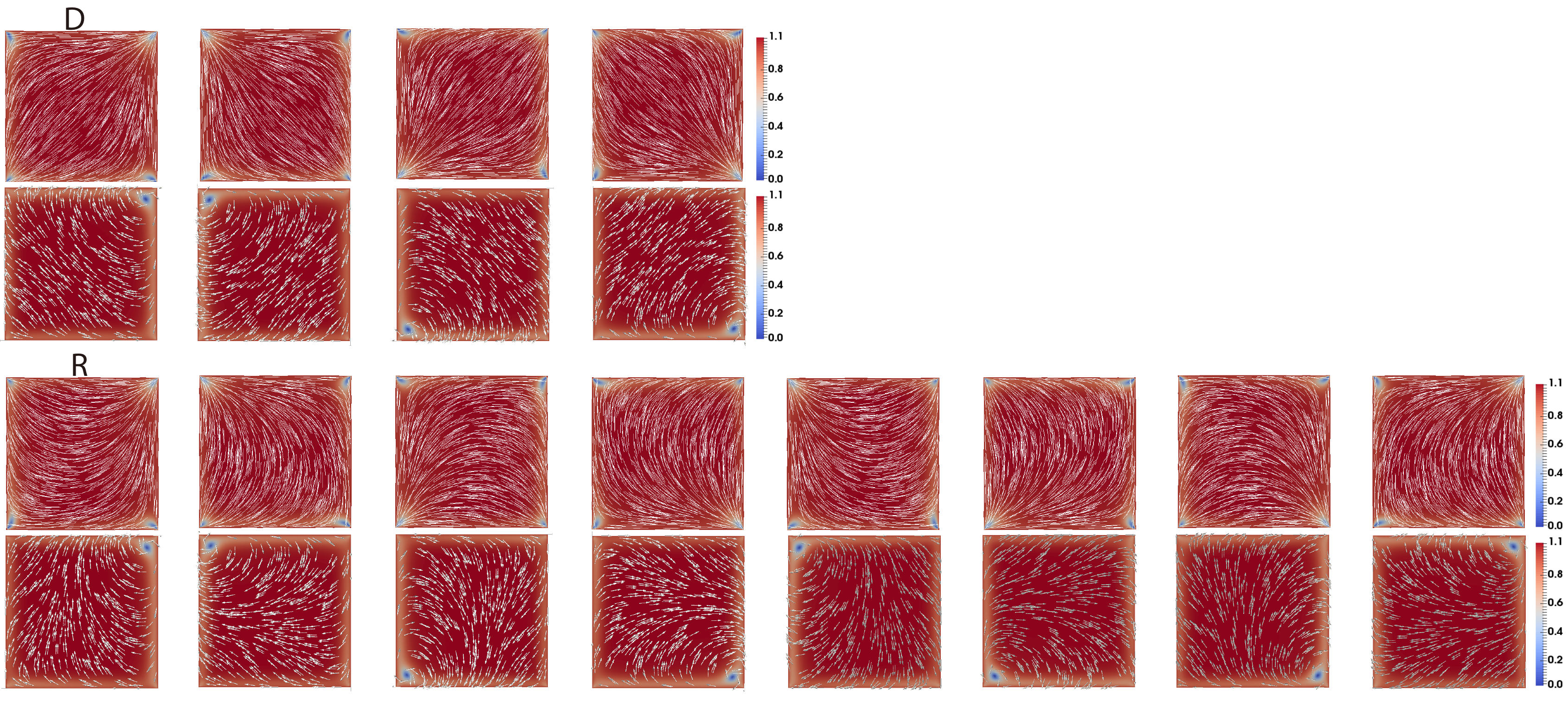}
        \caption{The plots of four $D$ and eight $R$ solutions with $\ell=10^{-3}$ and $c=-0.25$.}
        \label{fig:neg25_D_R}
\end{figure}

In Figure~\ref{fig:bifurcation_diagram_neg25}, we explore the solution landscape as a function of $\ell$, for $c=-0.25$. There are striking novelties here. For $\ell$ large, we observe the unique $Ring$ branch, which is globally stable for large $\ell$, and exists for all $\ell >0$. The $Ring$ branch loses stability as $\ell$ decreases. We note that the $Ring$ profile for small $\ell$ and $c=-0.25$, is different from its counterpart for $c=0.25$. This is essentially because $\n$ and $\M$ tend to be perpendicular in the square interior, since $c<0$. In particular, the $\Q$-solution in the $Ring$ branch adopts a hyperbolic-like central nematic defect structure, in sharp contrast to the vortex structure for $c=0.25$. The $\M$-profile has an interior magnetic vortex because of the topologically non-trivial Dirichlet conditions, as explained above.
As $\ell$ decreases, the $Ring$ branch bifurcates into \emph{$4$ stable D solutions}. This is notably different from the $c\geq 0$ case. Informally speaking, the symmetry between the splay vertices is broken in the nematic $D$ solution, rendering $4$ different $D$ solutions. One splay vertex is more asymmetric than the other splay vertex, and the corresponding $\M$-profile orients perpendicular to the $D$-director, with the magnetic vortex localised near the asymmetric splay vertex.
In the same vein, when $\ell$ is small enough, we find $8$ stable $R$ solution branches as seen in Figure \ref{fig:neg25_D_R}. The reasoning is the same as for the $D$ solutions. The symmetry between the splay vertices is broken for the $R$ solutions, with one splay vertex being 
more defective/asymmetric than the other splay vertex. Hence, there are $8$ $R$ solutions for the $\Q$-solution profile. The corresponding $\M$-profiles orient perpendicular to the nematic director 
 and the magnetic vortex localises near the more asymmetric splay vertex.

Additionally for small $\ell$, we find two stable $Peppa_{in}$ and two $Peppa_{out}$ solutions branches, with pairs of stable interior $+1/2$-nematic point defects. The $\M$-profiles for $Peppa_{in}$ ($Peppa_{out}$) have $\M$ pointing into (out of) the interior magnetic vortex, motivating the choice of the subscripts $in$ and $out$ respectively. The case of negative $c$ illustrates how we can use nemato-magnetic coupling to break symmetry, increase the multiplicity of stable solutions (for small $\ell$) and generate exotic permutations of defect profiles in $\Q$ and $\M$, all of which offer new prospects for engineered multistability.
\subsection{The $\ell\to 0$ limit.}

In this section, we study the asymptotics of minimizers of (\ref{energy}) in the $\ell \to 0$ limit, which is relevant for macroscopic domains, on the length scale of microns or larger. Recall that for $\ell_1 = \ell_2 = \ell$, $\xi=1$, the dimensionless free energy of this NLC (nematic liquid crystal)-MNP coupled system is given by: \begin{align} \label{energy2}
\mathcal{F}[\Q,\M]=&\int_\Omega\frac{1}{4}\left\{\ell |\nabla\Q|^2+\frac{1}{4}|\Q|^4-|\Q|^2\right\}\,\mathrm{dA}\nonumber\\
&+\int_\Omega \frac{1}{2}\left\{\ell|\nabla\M|^2+\frac{1}{2}|\M|^4-|\M|^2\right\}\,\mathrm{dA}\nonumber\\
&-\int_\Omega\frac{c}{2}\{Q_{11}(M_1^2-M_2^2)+2Q_{12}M_1M_2\}\,\mathrm{dA}.
\end{align}
In a 2D framework, we can parameterize $\Q$ and $\M$ as:
\begin{gather}
Q_{11}=S\cos(2\theta),\quad Q_{12}=S\sin(2\theta),\\
M_1=R\cos(\phi),\quad M_2=R\sin(\phi),
\end{gather}
where $|\Q|^2 = 2S^2$ and $|\M| = R$, and $\theta$, $\phi$ are orientation angles for $\n$ and $\M$ respectively.
Substituting the parameterization above into (\ref{energy2}), we obtain,
\begin{align}
\frac{1}{\ell} \mathcal{F}[S,R,\theta,\phi]=&  \int_\Omega \left\{\frac{1}{2}|\nabla S|^2+2S^2|\nabla\theta|^2\right\}\,\mathrm{dA}\nonumber\\
&+\int_\Omega\left\{\frac{1}{2}|\nabla R|^2+\frac{1}{2}R^2|\nabla\phi|^2\right\}\,\mathrm{dA}\nonumber\\
&+\frac{1}{\ell}\int_\Omega\left\{\frac{1}{4}S^4-\frac{1}{2}S^2+\frac{1}{4}R^4-\frac{1}{2}R^2 \right\}\,\mathrm{dA}\nonumber\\
&-\int_\Omega\frac{c}{2 \ell}SR^2\cos(2(\theta-\phi))\,\mathrm{dA}.	
\end{align}
Heuristically, the coupling energy determines the preferred relative orientation of $\n$ and $\M$. 
If $c>0$, the last term is minimized when
\begin{gather} 
\theta = \phi+\pi k,\quad k\in\mathbb{Z} \label{eq:positive_c}	
\end{gather}
i.e., when the director angle $\theta$ and the magnetization angle, $\phi$, differ by a multiple of $\pi$ so that $(\n \cdot \M) = \pm 1$. In particular, the coupling energy does not distinguish between $\M$ and $-\M$ and the $|\nabla \M|^2$ - energetic term penalises such arbitrary rotations. If $c<0$, the coupling energy is minimized when
\begin{gather}
\theta = \phi+\left(\frac{2k+1}{2}\right)\pi ,\quad k\in\mathbb{Z}	\label{eq:negative_c}
\end{gather}
i.e., for $\n \cdot \M = 0$.

Informally speaking, as $\ell \to 0$, minimizers of (\ref{energy2}) converge to appropriately defined minimizers of the bulk potential
\begin{align}
    f(S, R, \theta, \phi) =& \left(\frac{1}{4}S^4-\frac{1}{2}S^2 \right) + \left( \frac{1}{4}R^4-\frac{1}{2}R^2 \right)\nonumber\\
    &- \frac{c}{2}SR^2\cos(2(\theta-\phi)).
\end{align}
More precisely, in \cite{dalby21}, the authors compute the minimizers, $(S_c, R_c)$, of the bulk potential and show that
\begin{eqnarray}
    \label{eq:bulk_potential}
     S_c = &&\left( \frac{|c|}{4} + \sqrt{ \frac{c^2}{16} - \frac{1}{27}\left(1 + \frac{c^2}{2} \right)^3 } \right)^{1/3} + \nonumber \\ && + \left( \frac{|c|}{4} - \sqrt{ \frac{c^2}{16} - \frac{1}{27}\left(1 + \frac{c^2}{2} \right)^3 } \right)^{1/3};
     \nonumber \\
      R_c = &&\sqrt{ |c|S_c + 1}.
\end{eqnarray}
As $\ell \to 0$, for a fixed $c$, minimizers of (\ref{energy2}) converge (in an appropriately defined sense) to $\left( \Q^*, \M^* \right)$, where 
$|\Q^*| = \sqrt{2} S_c$, $|\M^*|  = R_c$ almost everywhere away from the polygon edges. The corresponding orientation angles, $\theta^*$ and $\phi^*$ are solutions of the Laplace equation
\begin{eqnarray}
&& \Delta \phi  = 0, \label{eq:l0}
\end{eqnarray} and $\theta^*$ and $\phi^*$ are related by (\ref{eq:positive_c}) for $c>0$, respectively (\ref{eq:negative_c}) for $c<0$, away from the polygon edges.

We can illustrate these concepts by considering the diagonal solutions in Figure~\ref{fig:25_D_R}, and the corresponding $\M$-profiles with domain walls along the square diagonals. For $c>0$ and small $\ell$, the preceding discussion suggests that $\theta$ and $\phi$ only differ by a multiple of $\pi$ in the interior. Let $c=0.25$ and consider one of the $D$ solutions. The corresponding boundary conditions for $\theta$ are\begin{equation}
\theta = \left\{  
\begin{array}{ll}
\frac{\pi}{2}, & x = \pm 0.5\\
0, & y= \pm 0.5
\end{array}
\right.
\end{equation}
However, this does not agree with the boundary conditions for $\phi$, which are fixed by 
(\ref{eq:square_bcs}) i.e.,
\begin{eqnarray} \label{eq:phi}
&& \phi = 0 ~ \textrm{on $y=0.5$;} ~ \phi = \pi ~\textrm{on $y=-0.5$}; \nonumber \\ 
&& \phi = \frac{\pi}{2}~ \textrm{on $x=-0.5$}; ~ \phi = \frac{3 \pi}{2}~ \textrm{on $x = 0.5$.}
\end{eqnarray}
Comparing the boundary conditions for $\theta$, for this $D$ solution, and $\phi$ above, along with the constraints imposed by (\ref{eq:positive_c}), we deduce that $\theta \approx \phi$ for $y \geq x$, and $\phi \approx \theta + \pi$ for $y < x$. Hence, there is a $\pi$-wall in the corresponding $\M$-profile along $x=y$ (see Figure~\ref{fig:25_D_R}). Analogous comments apply to the second $D$ solution (the second column in the first two rows of Figure~\ref{fig:25_D_R}) where we observe a $\pi$-wall in the $\M$-profile, along $y = -x$, such that $\phi$ flips by $\pi$-radians across the wall. $\M \neq 0$ on either side of the $\pi$-wall in these figures, so that these domain walls separate ordered magnetic polydomains. We refer to such $\pi$-walls as \emph{domain walls} in the subsequent discussion.

In Figure~\ref{fig:25_D_R}, there are $4$ stable $R$ solutions, labelled by say $R1 \ldots R4$. These rotated states can be defined by their boundary conditions e.g.,
\begin{eqnarray}
\label{eq:Rdomainwalls}
&& R1: \quad \theta (x, \pm 0.5) = 0; ~ \theta (-0.5, y) = \frac{3\pi}{2};~ \theta (0.5, y) = \frac{\pi}{2}, \nonumber \\
&& R2: \quad \theta (\pm 0.5, y) = \frac{\pi}{2}; ~ \theta (x, -0.5) = 0;~ \theta (x,0.5) = \pi, \nonumber \\
&& R3: \quad \theta (x, \pm 0.5) = 0; ~ \theta (-0.5, y) = \frac{\pi}{2};~ \theta (0.5, y) = \frac{3\pi}{2}, \nonumber \\
&& R4: \quad \theta (\pm 0.5, y) = \frac{\pi}{2}; ~ \theta (x, -0.5) = \pi;~ \theta (x, 0.5) = 0.\nonumber\\
\end{eqnarray}
These boundary conditions are incompatible with the boundary conditions for $\phi$ in (\ref{eq:square_bcs}). In the $\ell \to 0$ limit, with fixed $c>0$, we need $\theta$ and $\phi$ to differ by a multiple of $\pi$ almost everywhere. Comparing (\ref{eq:square_bcs}) 
  with the above, we deduce that the $\M$-profile corresponding to $R1$, has a domain wall near the edge $y=0.5$ i.e., $\phi$ rotates from $\phi=0$ to $\phi=\pi$ across a domain wall parallel to $y=0.5$, as can be clearly seen from the first column of the fourth row in Figure~\ref{fig:25_D_R}. In other words, $\theta \approx \phi$ for $y < 0.5$ and $\phi \approx \theta + \pi$ on $y=0.5$. Analogous remarks apply to the $\M$- profiles corresponding to $R2 \ldots R4$, where we observe domain walls along one of the square edges, such that $\theta \approx \phi$ on one side of the wall, and $\left| \theta - \phi \right| = \pi$ on the other side that contains the square edge in question. 

The $Peppa$ solution branch for positive coupling, is an example of the nematic profile being tailored by the magnetization profile. The boundary conditions for $\phi$ are fixed in (\ref{eq:square_bcs}) but the boundary conditions for $\theta$ are not fixed by (\ref{eq:square_bcs}), except that $2\theta$ is a multiple of $2\pi$ on $y=\pm 0.5$, and that $2\theta$ is an odd multiple of $\pi$ on $x=\pm 0.5$. In other words, $\theta$ can also assume the topologically non-trivial boundary conditions satisfied by $\phi$, and this is indeed observed in the $Peppa$-branch, for which the corresponding nematic director rotates by $2\pi$-radians along the boundary. The $2\pi$-rotation around the square perimeter necessarily means that $\n$ must have interior topological defects, with total charge of $+1$. For topological and energetic reasons, the $+1$-defect splits into two non-orientable $+1/2$-nematic defects in the interior, conserving the total topological charge. This is allowed in the reduced LdG framework, since the $\Q$-tensor includes non-orientable director fields, outside the scope of a vector field description. By contrast, the corresponding $\M$-profile has a single interior $+1$-vortex due to orientability constraints.

To summarize, for small $\ell$ and $c>0$, the $D$ and $R$ solution branches illustrate that the nematic profile can generate domains walls in the $\M$-profile, and the $Peppa$ solution branch demonstrates how the topologically non-trivial $\M$-profile can stabilise interior nematic point defects. The story with negative $c$ is more complex and fascinating, as we describe below.

We consider the diagonal solutions in Figure~\ref{fig:neg25_D_R} for $c=-0.25$. Consider $D1$ such that the nematic director, $\n$, is aligned along the square diagonal $y=x$. The corresponding $\M$ tends to be perpendicular to $\n$ in the interior, so that $\phi \approx \frac{3\pi}{4}$ or $\phi \approx \frac{-\pi}{4}$ along $y=x$. Further, the negative coupling breaks the symmetry between the two diagonally opposite splay vertices at $(0.5,0.5)$ and $(-0.5,-0.5)$. In the first column of the first row, the splay vertex at $(0.5,0.5)$ is ``more defective" than the second splay vertex, in the sense that $|\Q|(0.5,0.5) < |\Q|(-0.5, -0.5)$, and $\phi \approx \frac{3\pi}{4}$ along $y=x$ for the corresponding $\M$-profile, in the first column of the second row. Similarly, in the third columns of the first and second rows, the splay vertex at $(-0.5, -0.5)$, of $D1$ solution is more defective than the splay vertex at $(0.5,0.5)$, and $\phi \approx \frac{-\pi}{4}$ along $y=x$, for the corresponding $\M$-profile. Analogous remarks apply to the $D2$ solution with two splay vertices at $(-0.5, 0.5)$ and $(0.5, -0.5)$ respectively, with $\phi \approx \frac{\pi}{4}$ or $\phi \approx \frac{5 \pi}{4}$ along $y= -x$. The same reasoning applies to the $8$ rotated solutions in the third and fourth rows of Figure~\ref{fig:neg25_D_R}, for $c=-0.25$. Each of the $4$ rotated solutions for the $\Q$-profile is distinguished by two splay defects along a square edge. The negative coupling breaks the symmetry between the splay vertices so that one vertex is ``more asymmetric" than the other. This doubles the number of admissible rotated solutions. For each rotated solution, $\theta \approx 0$, $\theta \approx \pi$ or $\theta \approx \frac{\pi}{2}$, $\theta \approx \frac{3 \pi}{2}$ at the square centre. Each possibility generates two possibilities for $\phi$ at the square centre, for the corresponding $\M$-profiles in the fourth row of Figure~\ref{fig:neg25_D_R}. For example, $\phi \approx \frac{\pi}{2}$ or $\phi \approx \frac{3\pi}{2}$ (for $\theta \approx 0$ or $\theta \approx \pi$ near the centre) at the square centre, for the $\M$-profile, since negative $c$ coerces $\theta$ and $\phi$ to differ by an odd multiple of $\frac{\pi}{2}.$  These heuristic arguments corroborate the existence of $8$ rotated $(\Q, \M)$-stable solution profiles for small $\ell$, with $c = -0.25$.

Additionally, we find the $Peppa$ solution branches with stable interior nematic defects, as with positive $c$. For $c<0$, $\theta$ and $\phi$ tend to differ by an odd multiple of $\frac{\pi}{2}$ in the square interior, as $\ell \to 0$. In particular, this implies two choices for $\phi$ in the square interior, resulting in the $Peppa_{in}$ and $Peppa_{out}$ branches. There are two $Peppa_{in}$ solution branches, since the nematic defect pair can align along one of two square diagonals. Similarly, there are two $Peppa_{out}$ solution branches by the same reasoning as above. The case of negative coupling strongly enhances multistability for small $\ell$, effectively doubling the number of admissible stable states compared to positive coupling (compare Figures~\ref{fig:bifurcation_diagram_25} and \ref{fig:bifurcation_diagram_neg25}). We do not observe domain walls in $\M$ for negative coupling, rather we observe magnetic vortices at the square vertices for negative coupling.
 These corner defects may act as distinguished sites/binding sites for devices based on such NLC-MNP systems.


\subsection{The $\ell\to\infty$ limit.}
The $\ell \to \infty$ limit is relevant for small nano-scale domains. Mathematically, this limit is much simpler than the $\ell \to 0$ limit, since we lose the nemato-magnetic coupling in this limit. Referring to \cite{FangLidong2020Ssat}, the leading order equations, in this limit, are:
\begin{eqnarray}
&& \Delta \Q = \mathbf{0},
 \nonumber \\ && \Delta \M = \mathbf{0},
  \label{eq:linfty}
 \end{eqnarray}
 subject to the Dirichlet conditions (\ref{eq:square_bcs}). The limiting solution is unique. It is straightforward to recover the WORS for the $\Q$-profile, and to show that there is a magnetic vortex of degree $+1$ at the square centre (with $\M (0,0) = 0$), for the $\M$-profile, in this limit. This is precisely the solution along the $Ring$ branch for large $\ell$, in the bifurcation diagrams Figures~\ref{fig:bifurcation_diagram_25} and \ref{fig:bifurcation_diagram_neg25}, which is the unique energy minimizer in this limit.
 
 Following the methods in \cite{FangLidong2020Ssat}, the limiting solution, $\left(\Q^\infty, \M^\infty \right)$ of (\ref{eq:linfty}) is an excellent approximation to the solutions, $\left(\Q^{\ell}, \M^{\ell} \right)$ of (\ref{EL1})-(\ref{EL4}), for fixed $c$, subject to the same boundary conditions, for $\ell$ large enough i.e., $\left| \left(\Q^{\ell}, \M^{\ell} \right) - \left(\Q^{\infty}, \M^{\infty} \right)\right|^2 \sim \frac{1}{\ell^2}$. 
 
\begin{figure}
		\centering
        \includegraphics[width=0.5\columnwidth]{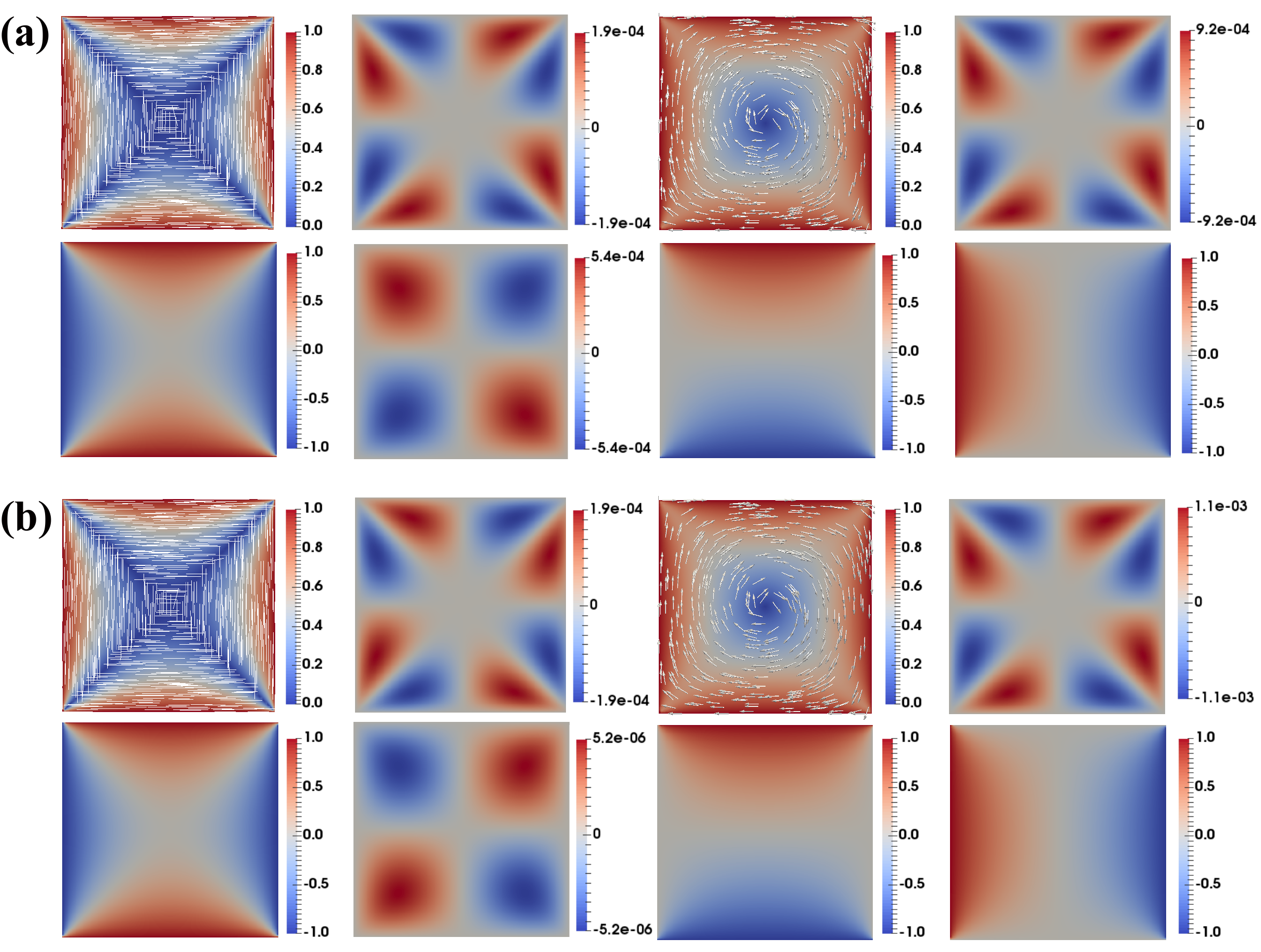}
        \caption{(a) and (b) are relative plots of the $Ring$ solution at $c = 0.25$ and $c = -0.25$, respectively. The plots in the first row of (a) and (b) from left to right are: $\mathbf{Q}^1$; the difference $Q_{11}^1Q_{12}^{100}-Q_{11}^{100}Q_{12}^1=$ $S^1S^{100}\sin(2\theta^{100}-2\theta^1)$; $\mathbf{M^1}$ and; the difference $M_1^1M_2^{100}-M_1^{100}M_2^1 = R^1R^{100}\sin(\phi^{100}-\phi^{1})$. In the $\mathbf{Q}^1$ plot, the corresponding vector $\mathbf{n}^1$ in \eqref{eq:n} is represented by white lines and the order parameter $|\mathbf{Q}^1|/\sqrt{2} =$ $\sqrt{{Q_{11}^2}^2 + {Q_{12}^1}^2}$ is labelled by the colour chart. In the $\mathbf{M}^1$ plot, $|\M^1|$ is labelled by the colour chart, and the white arrows describe the magnetic orientation $(M_1^1,M_2^1)/ |\M^1|$ for $|\M^1|\neq 0$. The plots in the second row of (a) and (b) from left to right are: $Q_{11}^1$; $Q_{12}^1$; $M_1^1$ and; $M_2^1$.}
        \label{fig:Ring_25_neg25}
\end{figure}

 However, the unique limiting solution $(\Q^\infty, \M^\infty)$ remains an excellent approximation to the $Ring$ solution, even for values of $\ell$ as small as unity as we show below. We demonstrate this by comparing two solutions along the $Ring$ branch, for $\ell =1$ and $\ell=100$, denoted by $\left(\mathbf{Q}^1,\mathbf{M}^1 \right)$ and $\left(\mathbf{Q}^{100},\mathbf{M}^{100}\right)$ respectively. The solution, $\left(\mathbf{Q}^{100},\mathbf{M}^{100} \right)$ is effectively identical to the limiting solution $\left(\Q^{\infty}, \M^{\infty} \right)$ described above. Let 
 \begin{align}
     (Q_{11}^1,Q_{12}^1) &= S^1(\cos2\theta^1,\sin2\theta^1), \\(Q_{11}^{100},Q_{12}^{100}) &= S^{100}(\cos2\theta^{100},\sin2\theta^{100}), \\
     (M_1^1,M_2^1) &= R^1(\cos\phi^1,\sin\phi^1), \\ (M_1^{100},M_2^{100}) &= R^{100}(\cos\phi^{100},\sin\phi^{100}).
 \end{align}
In Figure~\ref{fig:Ring_25_neg25}, we plot the differences between the orientation angles, $\sin(2\theta^{100}-2\theta^1)$ and $\sin(\phi^{100}-\phi^{1})$, and they are of the order of $10^{-4}$, from which we deduce that $(\Q^{\infty}, \M^{\infty})$ is a reliable approximation to $(\Q^{\ell}, \M^{\ell})$, along the $Ring$ branch for $\ell \geq 1$.

\section{Hexagons}

Next, we consider a NLC-MNP suspension on a 2D regular hexagon, 
subject to the Dirichlet conditions for $\Q$ and $\M$ in (\ref{BC1}) and (\ref{BC2}), for $N=6$ respectively.

The case of $c=0$ has been well studied in \cite{han2020reduced}. For large $\ell_1 = \ell_2 = \ell$ and $c=0$, there is a unique $Ring$ solution on the hexagon, for which the corresponding $\Q$ and $\M$ profiles have a single $+1$-vortex at the centre of the hexagon. This $Ring$ solution branch loses stability as $\ell$ decreases. In the limit of small $\ell$, with $c = 0$, there are at least $15$ different stable states, with topologically trivial boundary conditions i.e.,
\begin{gather}\label{deg_0}
\mathrm{deg}(\n_b,\partial\Omega)=0,
\end{gather}
which represents the Brouwer degree or winding number of $n_b$ considered as a map from $\partial\Omega$ into $S^1$.
$\n_b$ and $\Q_b$ are related by
\begin{gather}
\Q_b=
\begin{pmatrix}
Q_{11b} & Q_{12b} \\
Q_{12b} & - Q_{11b}\\	
\end{pmatrix}
 =  \left( 2\n_b \otimes \n_b - \mathbf{I} \right).
\end{gather}
These $15$ states are categorised by permutations of the vertex defects. There are $6$ vertices, two of which have $+1/3$-charge (referred to as \emph{splay} vertices) and four of which have $-1/6$-charge (referred to as \emph{bend} vertices). These 15 solutions are split into 3 rotationally invariant classes: (i) the 3 \textit{Para} states, where the splay defects are opposite each other; (ii) the 6 \textit{Meta} states, where the splay defects are separated by one vertex; and (iii) the 6 \textit{Ortho} states, where the splay defects are connected by an edge. In \cite{han2020reduced}, the authors show that there exist 3 bifurcation points  such that the \textit{Ortho}, \textit{Meta} and \textit{Para} states gain stability for $\ell<\ell_{Ortho}<\ell_{Meta}<\ell_{Para}$ respectively, for $c=0$.
\begin{figure}
    \begin{center}
    \includegraphics[width=0.5\columnwidth]{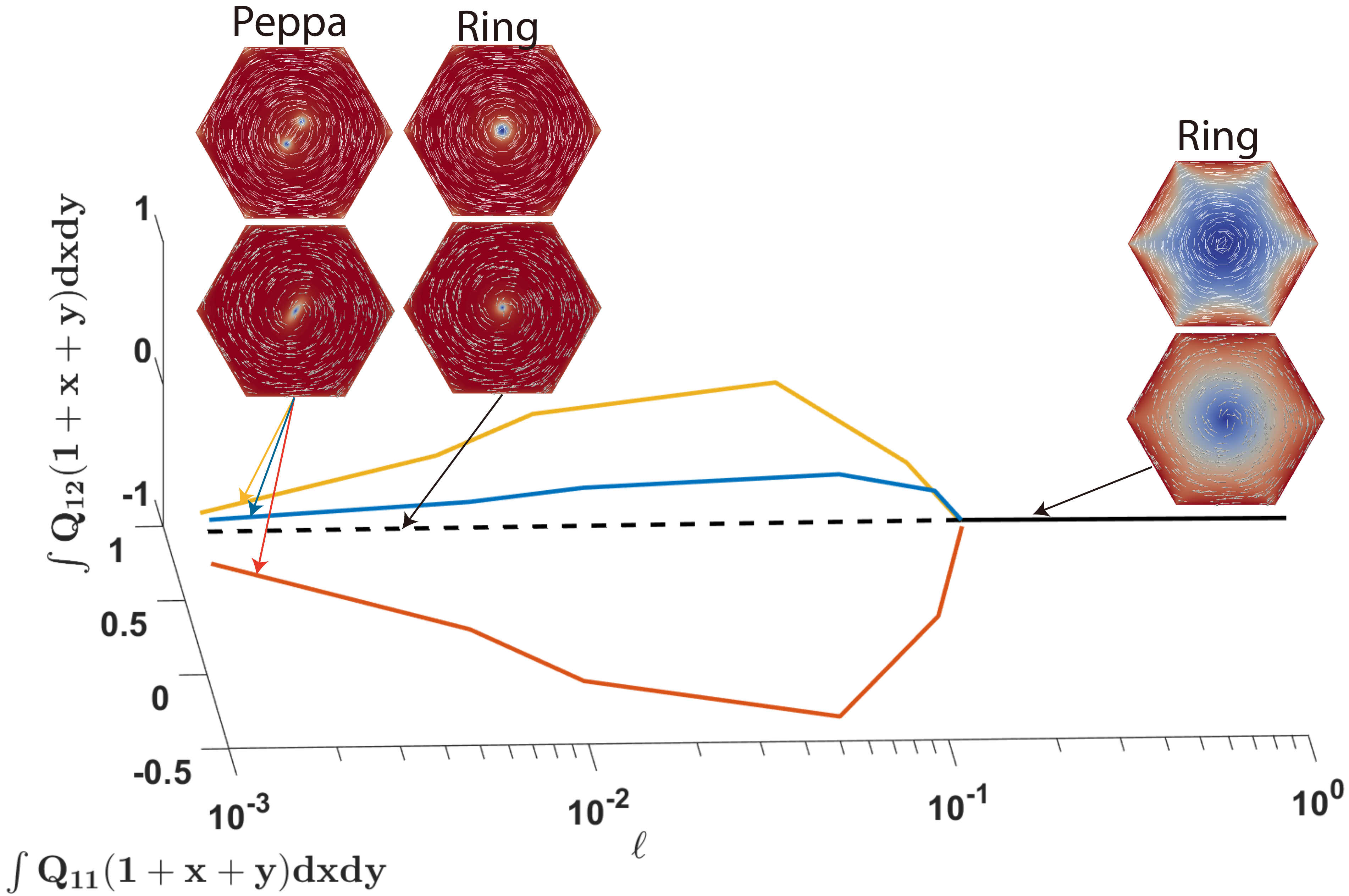}
    \caption{Bifurcation diagram for (\ref{energy}) on a hexagon domain with $c = 0.25$ plotting $\int Q_{11}\left(1+x+y\right)\textrm{d}x\textrm{d}y$ and $\int Q_{12}\left(1+x+y\right)\textrm{d}x\textrm{d}y$ versus $\ell$.}
    \label{fig:hexagon_bifurcation_diagram_25}
    \end{center}
\end{figure}

In Figure~\ref{fig:hexagon_bifurcation_diagram_25}, we track the different solution branches as a function of $\ell$ with $c=0.25$, using the $Ring$ solution, the \textit{Para}, \textit{Meta} and \textit{Ortho} states as initial conditions for the $\Q$-solver. The $Ring$ branch exists for all $\ell>0$, is unique and globally stable for $\ell$ large enough, but loses stability as $\ell$ decreases, as expected by analogy with the $c=0$ case. In contrast to a square domain, we lose the \textit{Para}, \textit{Meta}, \textit{Ortho} solutions and only recover three $Peppa$ solution branches in the $\ell \to 0$ limit. The three $Peppa$ solution branches are featured by a pair of stable interior $+1/2$-nematic defects aligned along one of the hexagon diagonals, near the center of the hexagon. There are three hexagon diagonals, and hence there are three $Peppa$ solution branches. The corresponding $\M$-profiles have a slightly smeared magnetic vortex along the line connecting the nematic defect pair. We will explore this in greater detail below, but magnetic domain walls connecting pairs of diagonally opposite vertices for the $Para$-nematic state on a regular hexagon, have greater length than their corresponding counterparts on a square domain. Magnetic domain walls for $Meta$-nematic states have lesser symmetry. Heuristically, this may explain the absence of magnetic domain walls in stable $(\Q, \M)$-profiles on a hexagon, with $c=0.25$. Equally, our numerical methods are not exhaustive, and we may have omitted certain solution branches e.g., high energy $\textit{Meta}$ and $\textit{Ortho}$ solution branches.
\begin{figure}
\centering
    \begin{subfigure}{0.74\columnwidth}
        \centering
        \includegraphics[width=\textwidth]{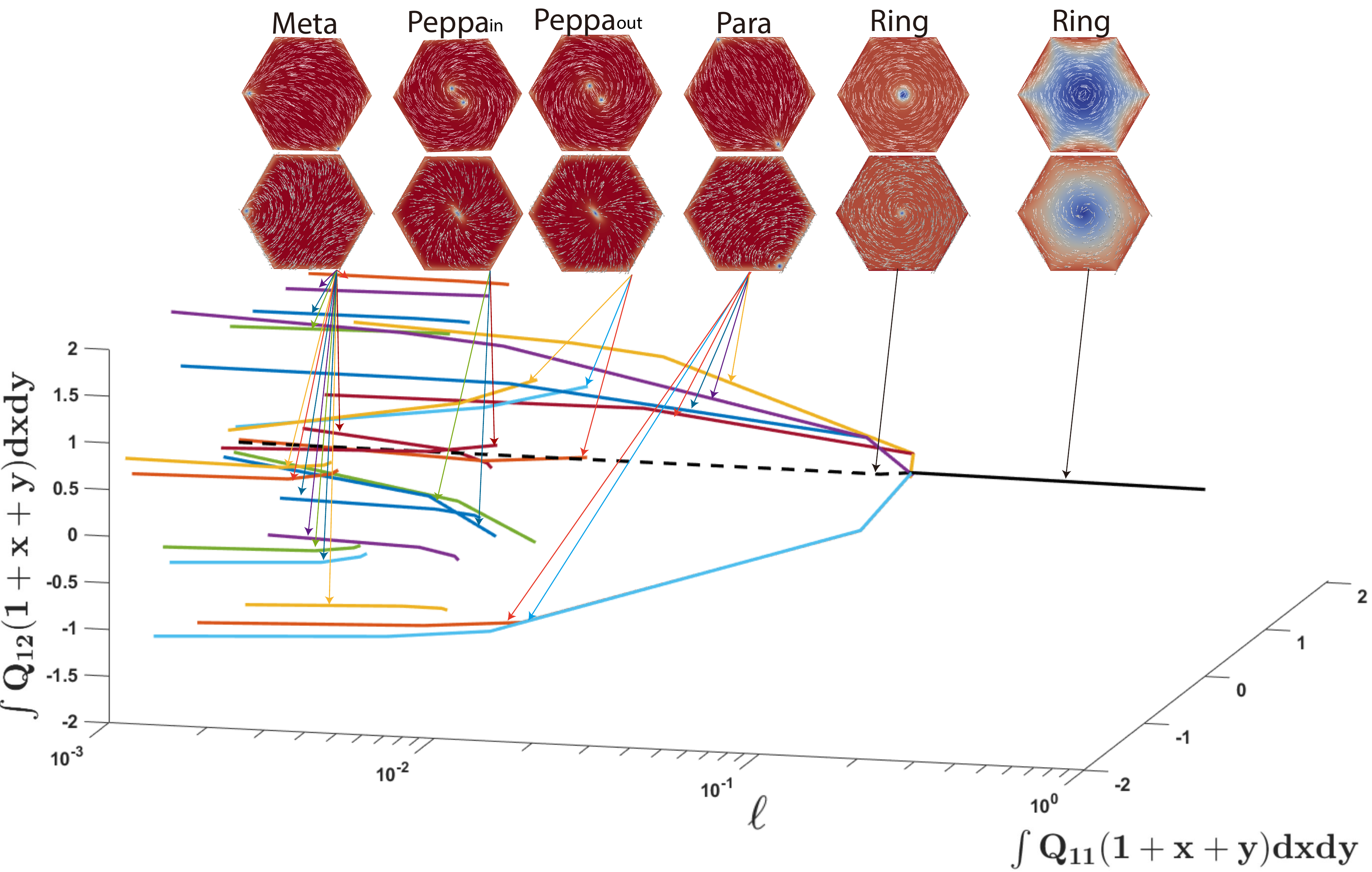}
    \end{subfigure}
    \begin{subfigure}{0.25\columnwidth}
        \centering
        \includegraphics[width=\textwidth]{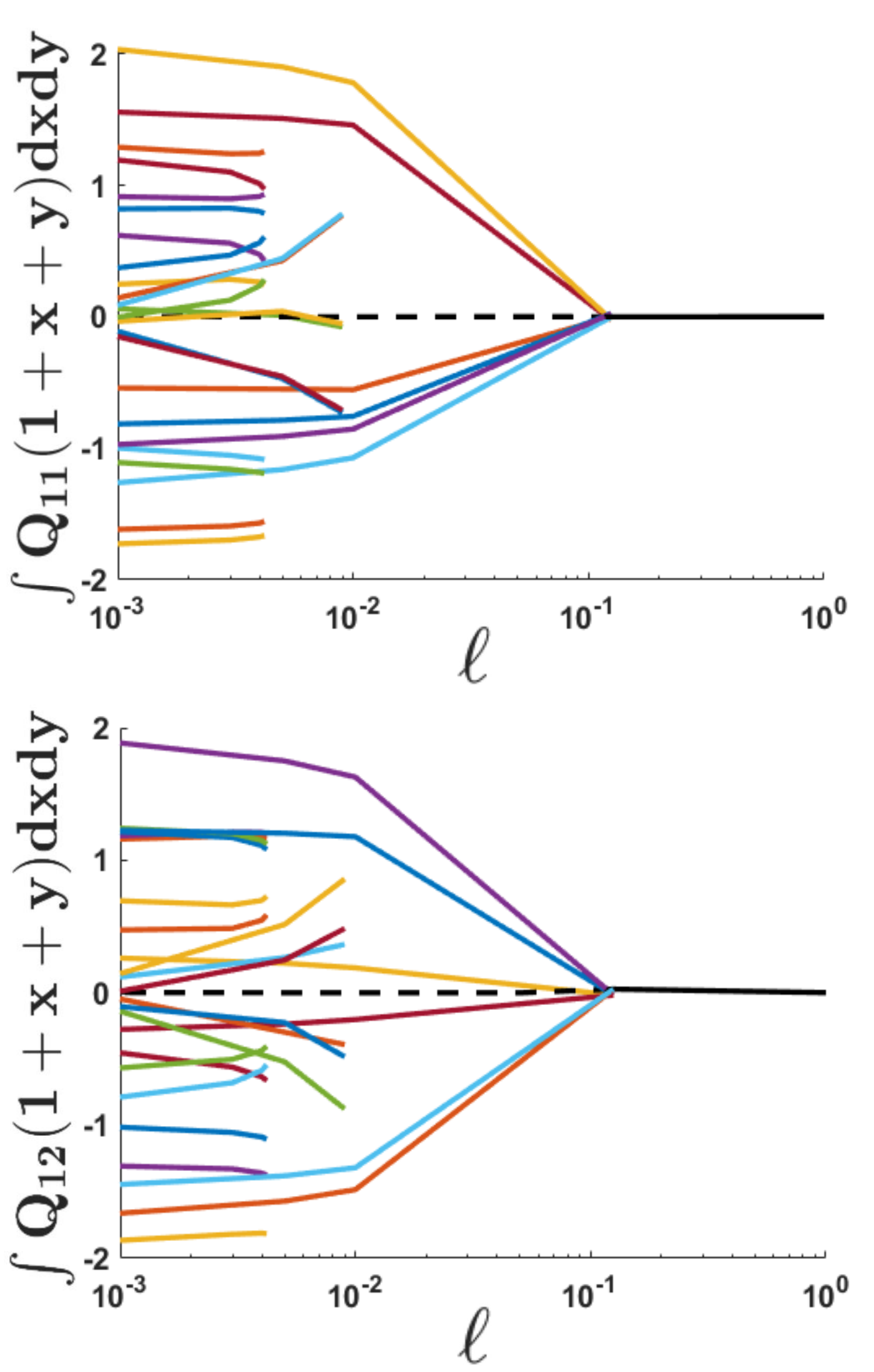}
    \end{subfigure}
        \caption{Bifurcation diagram for (\ref{energy}) on a hexagon domain with $c = - 0.25$. Left : plot of $\int Q_{11}\left(1+x+y\right)\textrm{d}x\textrm{d}y$, $\int Q_{12}\left(1+x+y\right)\textrm{d}x\textrm{d}y$ versus $\ell$; right: orthogonal 2D projections of the full 3D plot.}
        \label{fig:hexagon_bifurcation_diagram_neg25}
\end{figure}

\begin{figure}
	\centering
	\includegraphics[width=0.3\columnwidth]{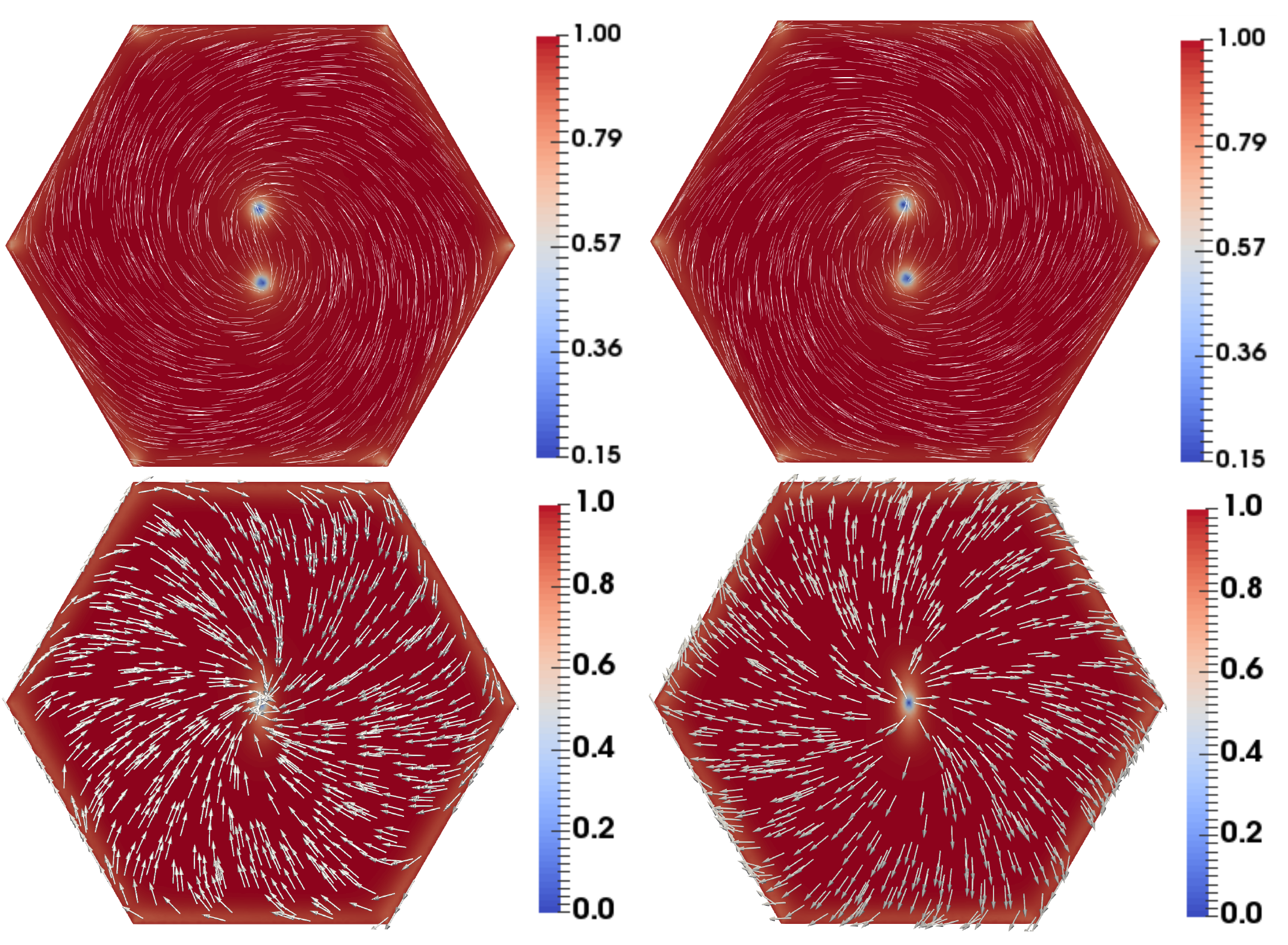}
	\caption{$Peppa_{in}$ and $Peppa_{out}$ solution profiles (from left to right) for $c=-0.1$, with $\xi=1$, and $\ell=5\times10^{-4}$. }
	\label{fig:spiral}
\end{figure}

In Figure~\ref{fig:hexagon_bifurcation_diagram_neg25}, we plot the solution landscape on the re-scaled hexagon, as a function of $\ell$, for $c=-0.25$. As before, we have a unique and globally stable $Ring$ solution branch for large $\ell$, which exists for all $\ell>0$ and loses stability as $\ell$ decreases. Here, as with a square domain, we effectively double the number of $Ortho$, $Meta$ and $Para$ states, since the symmetry between the splay vertices is broken. We numerically observe six $Para$, twelve $Meta$, twelve $Ortho$-nematic states. The corresponding $\M$-profiles are distinguished by the location of the magnetic vortex at one of the hexagon vertices (six possibilities) and the orientation of $\M$, since $\M$ is preferentially perpendicular to $\n$ in the hexagon interior for $c=-0.25$. For example, there are six $Para$ $\left(\Q, \M \right)$-states, corresponding to six possibilities for the location of the more asymmetric splay vertex. For $c=0$, there are six $Meta$ stable states for $\ell$ small enough, with two splay vertices separated by a vertex. For $c<0$, the symmetry between the splay vertices is broken and we obtain two $Meta$ states for each admissible \emph{splay vertex pair}, yielding a total of $12$ $Meta$ states. 
Analogous remarks apply to the $Ortho$ solution branch. As with the square, we also observe three $Peppa_{in}$ and $Peppa_{out}$ solution branches, with pairs of stable interior nematic defects along the three hexagon diagonals. The $\textit{in}$-branches refer to inwards-pointing $\M$-profiles, and $\textit{out}$-branches refer to outward-pointing $\M$-profiles from the central magnetic vortex. In Figure \ref{fig:spiral}, we plot a $Peppa_{in}$ and $Peppa_{out}$ solution (the $(\Q, \M)$ profiles), for $c=-0.1$ with $\xi=1$ and $\ell=5\times10^{-4}$.

\begin{figure}
	\centering
	\includegraphics[width=0.4\columnwidth]{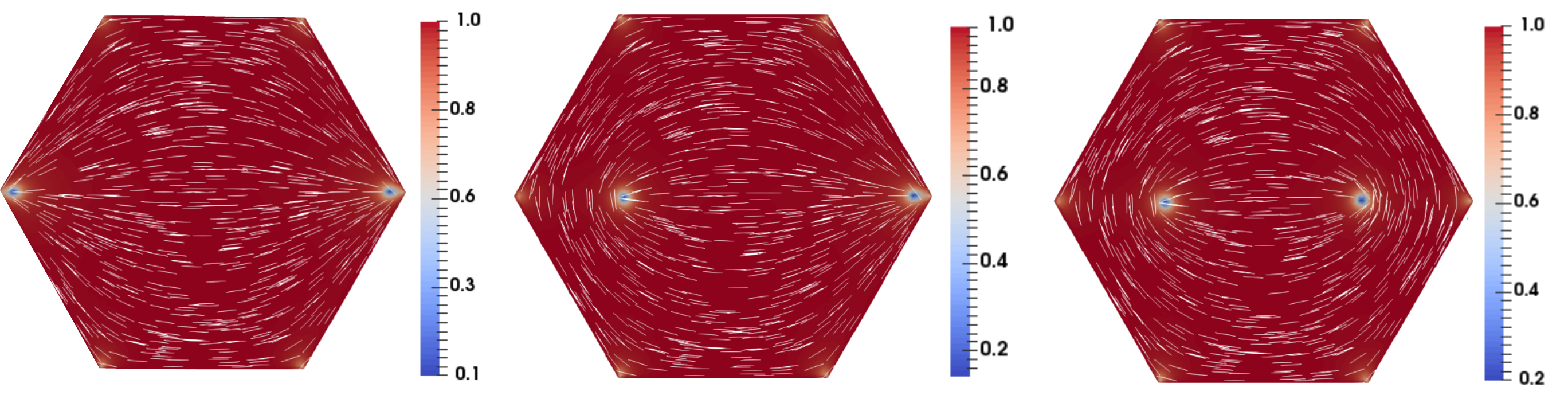}
	\caption{The nematic $\Q$-profile for small and positive coupling ($c=5,6,7\times10^{-3}$, respectively). The \textit{Para} solution for $c=0$ was taken as an initial guess with $\ell, \xi$ fixed. Here we see the \textit{Para} solutions transition into the \textit{Peppa} solution as $c$ increases.}
	\label{fig:defectPara6}
\end{figure}

To understand how the splay defects evolve in the $Para$ solution branch for $c>0$, we use the $Para$-nematic solution (for $c=0$) as an initial condition for small values of $c = 5,6,7 \times 10^{-3}$, $\ell = 10^{-4}$, $\xi=1$ to trace the $Para$ branch using continuation methods, see Figure~\ref{fig:defectPara6}. As we move from left to right i.e., from $c=5\times 10^{-3}$ to $c = 7 \times 10^{-3}$, it is clear that the defects detach from the splay vertices as $c$ increases, and migrate towards the hexagon interior, and align along one of the hexagon diagonals as $c\to1$. As $c$ increases, the interior nematic defects localise near the centre of the hexagon, yielding the $Peppa$ solution branches in Figure~\ref{fig:hexagon_bifurcation_diagram_25}. The $Peppa$ solution branches are clear examples of nematic profiles being tailored by the magnetic profile. Namely the central magnetic vortex coerces the creation of two $+1/2$-stable interior nematic defects, due to the positive nemato-magnetic coupling that favours co-alignment of $\n$ and $\M$. 

\begin{figure}
	\centering
	\includegraphics[width=0.4\columnwidth]{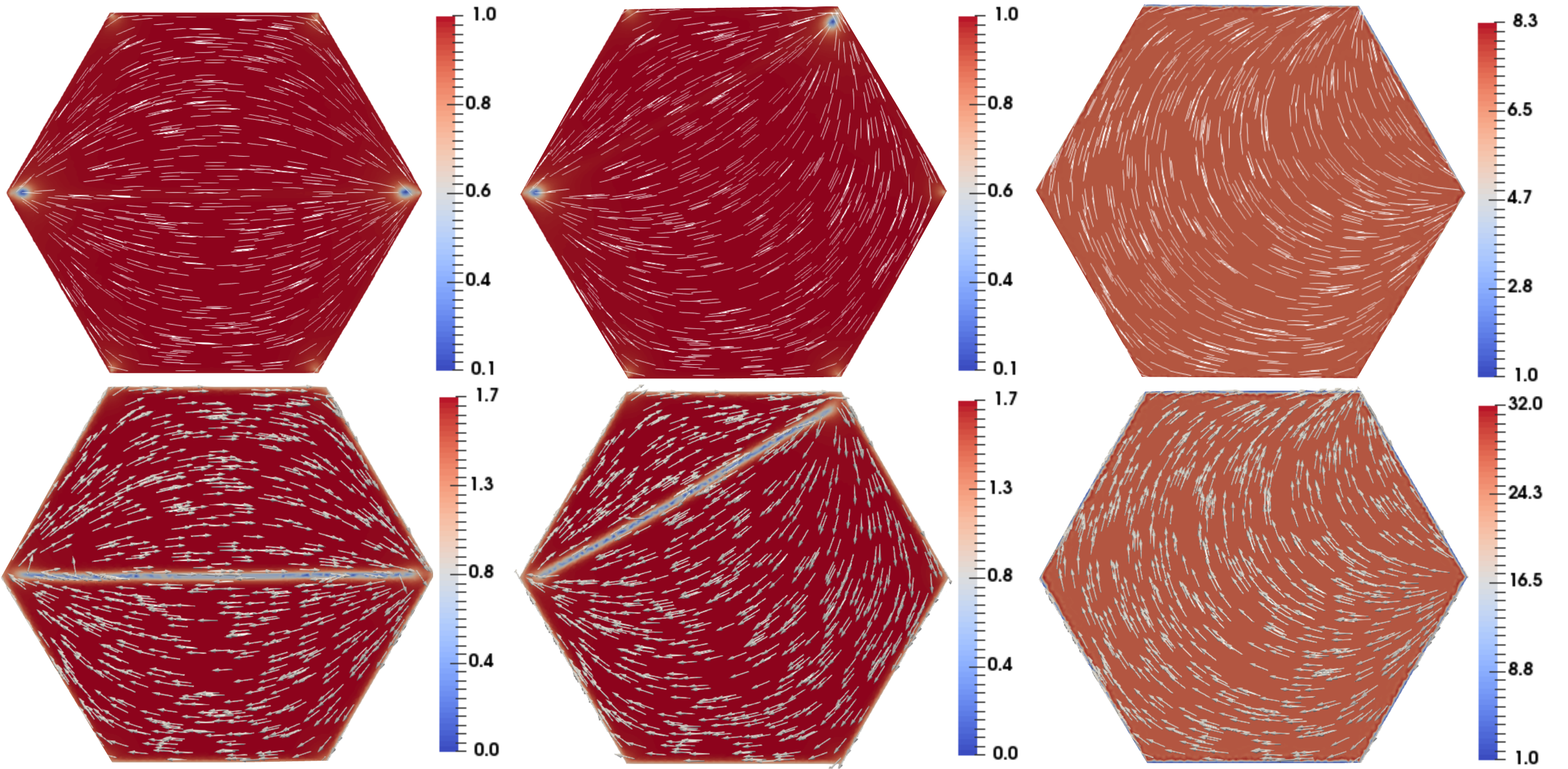}
    \caption{\textit{Para} and \textit{Meta} solution profiles for $c=0.02$ and the \textit{Ortho} solution profile for $c=1$, respectively with $\xi=0.01$ and $\ell=5\times 10^{-4}$.}\label{fig:smallXI}
\end{figure}

For a square domain, we observe diagonal ($D$) and rotated ($R$) solution branches for $\ell$ small enough, $c=0.25$, for which the corresponding $\M$-profile exhibits a domain wall, either along a square diagonal or along a square edge respectively. These domain walls are characterized by a sharp drop in $|\M|$ compared to the surrounding values. It is evident that these \emph{domain wall} $\M$-profiles are increasingly difficult to find in a hexagon for positive $c$, and in a pentagon as will be shown below. In the preceding simulations, $\xi=1$. We conjecture that smaller values of $\xi$ will coerce the $\Q$-profile to tailor the $\M$-profile for $c>0$ i.e., the $\M$-texture will be determined by $\n$, leading to the creation of domain walls in $\M$. A smaller value of $\xi$ suppresses the magnetic energy and hence, the nematic effects dominate in this regime. The domain walls are essentially a consequence of the topologically non-trivial boundary conditions for $\M$, so that $\n \cdot \M \approx 1$ on one side of the wall, and $\n \cdot \M \approx -1$ on the other side of the wall. 
In Figure~\ref{fig:smallXI}, we take $\xi=0.01$, $\ell = 5 \times 10^{-4}$, and use the $Para$, $Meta$ and $Ortho$-nematic solutions (for $c=0$) and the $\M$-solution with a central magnetic vortex (for $c=0$) as initial conditions. We do indeed recover the $Para$ solution for the $\Q$-profile with two defects pinned at a pair of diagonally opposite splay vertices, and the corresponding $\M$-profile has a clear domain wall along the diagonal connecting the splay vertices, for $c \leq 0.02$. Analogous remarks apply to $Meta$ solutions, for which the $\M$-profile has a distinct domain wall along the line connecting the two splay vertices. In other words, we can numerically find $Meta$ solutions for which the $\Q$-profile with two splay vertices (separated by a vertex) and $\M$ has an associated domain wall, for $0<c\leq 0.02$.
The $Ortho$ solutions are easier to find, with a short magnetic domain wall along the hexagon edge connecting the two adjacent splay vertices in the $Ortho$ $\Q$-solution. We find these $Ortho$ solutions by continuation methods for $c\leq 1$. 

We deduce that we can stabilise either interior nematic point defects or magnetic domain walls, depending on a judicious interplay of $\xi$ and $c$, and this interplay depends on $N$ - the number of sides of the regular polygon. A reasonable conjecture is that magnetic domain walls are observable for $\xi < \xi (N)$ and $0< c< c(N)$, for the boundary conditions in (\ref{BC1}) and (\ref{BC2}). We expect that $\xi(N)$ and $c(N)$ are decreasing functions of $N$ i.e. $\xi(N) \to 0, c(N) \to 0$ as $N\to \infty$, so that domain walls are increasingly difficult to find for coupled systems.

\section{Pentagons}

Next, we consider a regular pentagon with $N=5$, and study the solution landscape as a function of $\ell$, for positive $c$ and negative $c$ respectively ($c=0.25$ and $c=-0.25$ respectively). The case of $c=0$ has been well studied in \cite{han2020reduced}. For $c=0$ and $\ell$ large enough, there is a unique $Ring$ solution for the $\Q$-solution (with a central $+1$-nematic defect) and a unique $\M$-profile with a degree $+1$ central vortex. This $Ring$ branch is globally stable for $\ell$ large enough, exists for all $\ell>0$ and is unstable for $\ell$ small enough. For small $\ell$, there are at least $10$ different stable solutions (with $c=0$) for the $\Q$-solutions, for topologically trivial boundary conditions \eqref{deg_0}.
As with the hexagon, the tangent boundary conditions naturally create a mismatch in $\n_b$ at the pentagon vertices, so that the vertices are natural candidates for nematic defects. The different vertices are classified as ``splay" and ``bend" vertices, and there are two splay, and three bend vertices for topologically trivial boundary conditions. The 10 solutions are classified into 2 rotationally invariant classes: (i) the 5 $Meta$ states, where the splay vertices are separated by one vertex; and (ii) the 5 higher energy $Ortho$ states, where the splay vertices are connected by an edge. In \cite{han2020reduced}, the authors show that there exist at least 2 bifurcation points such that the $Meta$ and $Ortho$  states gain stability for $l<l_{Ortho}<l_{Meta}$ respectively.
\begin{figure}
    \centering
    \includegraphics[width=0.5\columnwidth]{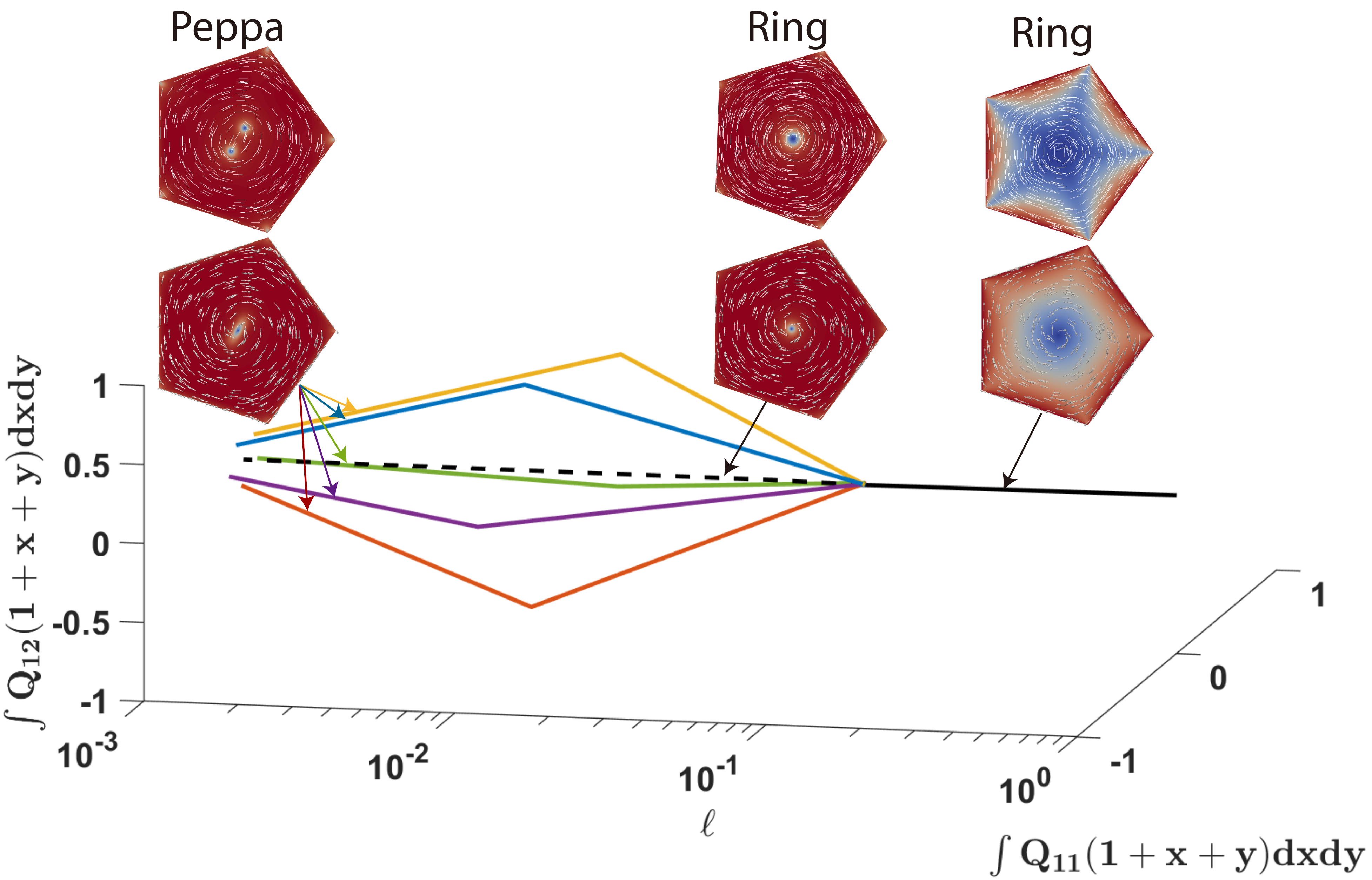}
    \caption{Bifurcation diagram for (\ref{energy}) on a pentagon domain with $c = 0.25$. The plot of $\int Q_{11}\left(1+x+y\right)\textrm{d}x\textrm{d}y$, $\int Q_{12}\left(1+x+y\right)\textrm{d}x\textrm{d}y$ versus $\ell$.}
    \label{fig:pentagon_bifurcation_diagram_25}
\end{figure}

\begin{figure}[t!]
    \centering
        \includegraphics[width=0.3\columnwidth]{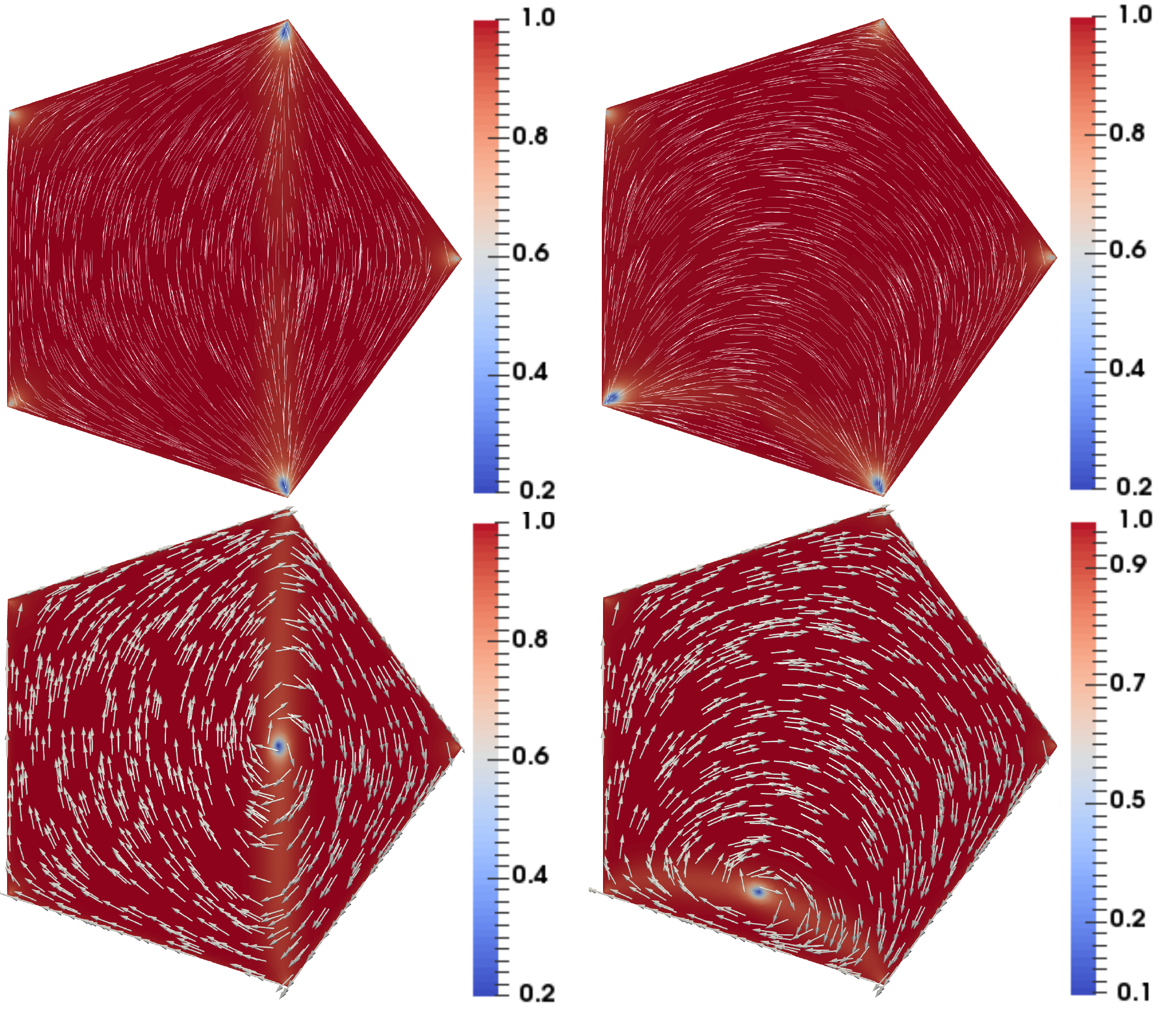}
        \caption{\textit{Meta} and \textit{Ortho} solution profiles (from left to right) for $c=0.05$, respectively. We take the \textit{Meta} and \textit{Ortho} solutions on a pentagon for $c=0$ as the initial guesses and fix $l=5\times10^{-4}$ and $\xi=1$. For $c$ small enough, the nematic profile is maintained, whereas for $c$ large, we get the pair of interior point defects. There is a reduction in $S$ along the polygon edges as $c\to1$.}
        \label{fig:pentaSmallc}
\end{figure}

In Figure~\ref{fig:pentagon_bifurcation_diagram_25}, we plot the bifurcation diagram for the $(\Q, \M)$-solutions as a function of $\ell$, for $c=0.25$. The qualitative features are similar to those for a hexagon, we lose the $Ortho$ and $Meta$ states and obtain five stable $Peppa$ solution branches for $\ell$-small enough, with two stable interior nematic defects. For each $Peppa$ branch, the nematic defect pair is localised near the centre of the pentagon, parallel to one of the pentagon edges. The magnetic profile retains the central vortex and the $Peppa$ solutions are again examples of nematic profiles tailored by the magnetic profile.
However, domain walls are easier to find in pentagons compared to hexagons. In Figure~\ref{fig:pentaSmallc}, we recover the $Meta$ and $Ortho$-nematic states in a pentagon, with $\xi=1$, $\ell = 5\times 10^{-4}$ with $c=0.05$, which is not observed in a hexagon. The corresponding $\M$-profiles exhibit domain walls (with reduced $|\M|$) along straight lines connecting the splay vertices.

\begin{figure}
\centering
    \begin{subfigure}{0.74\columnwidth}
        \centering
        \includegraphics[width = \textwidth]{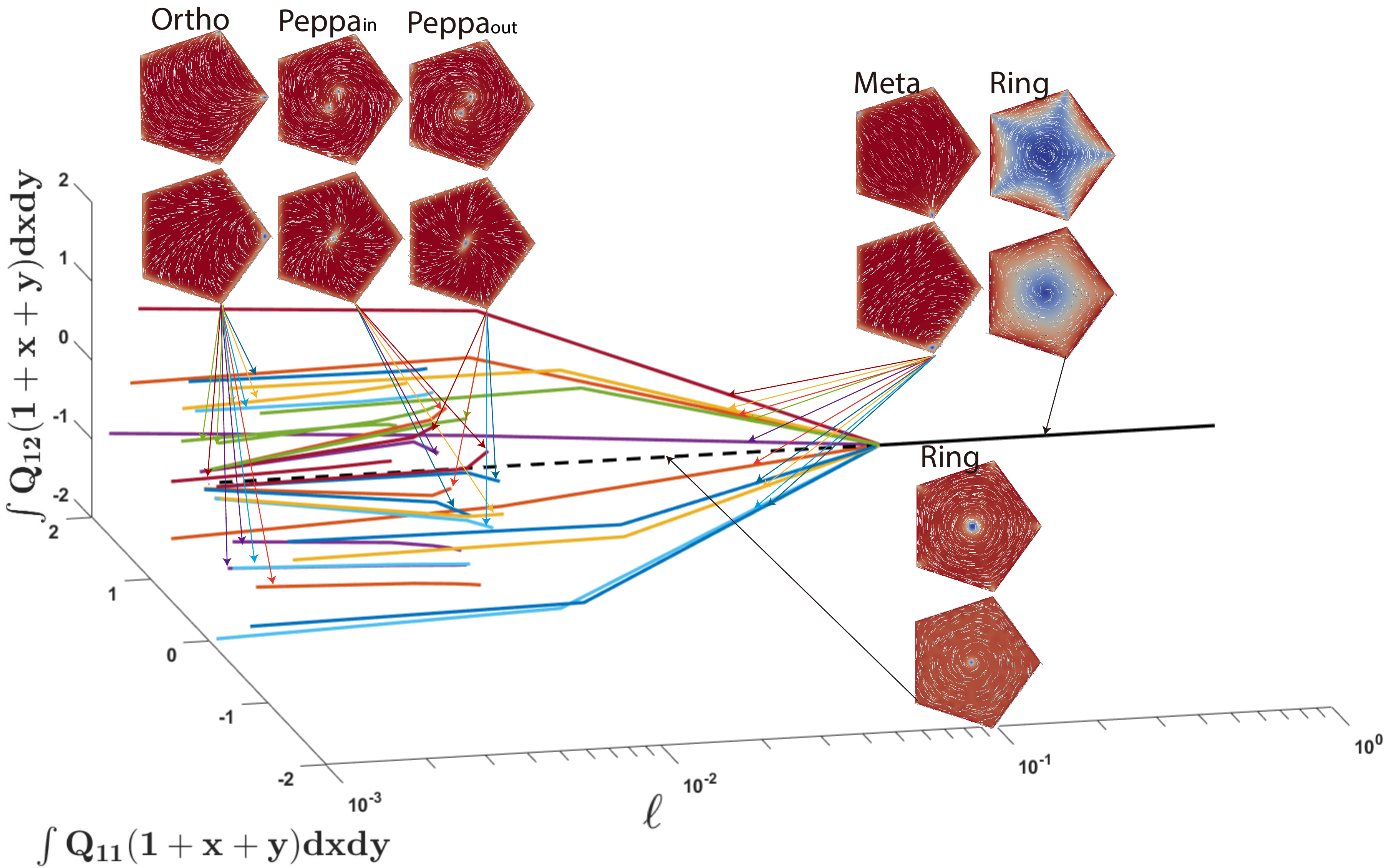}
    \end{subfigure}
    \begin{subfigure}{0.25\columnwidth}
        \centering
        \includegraphics[width=\textwidth]{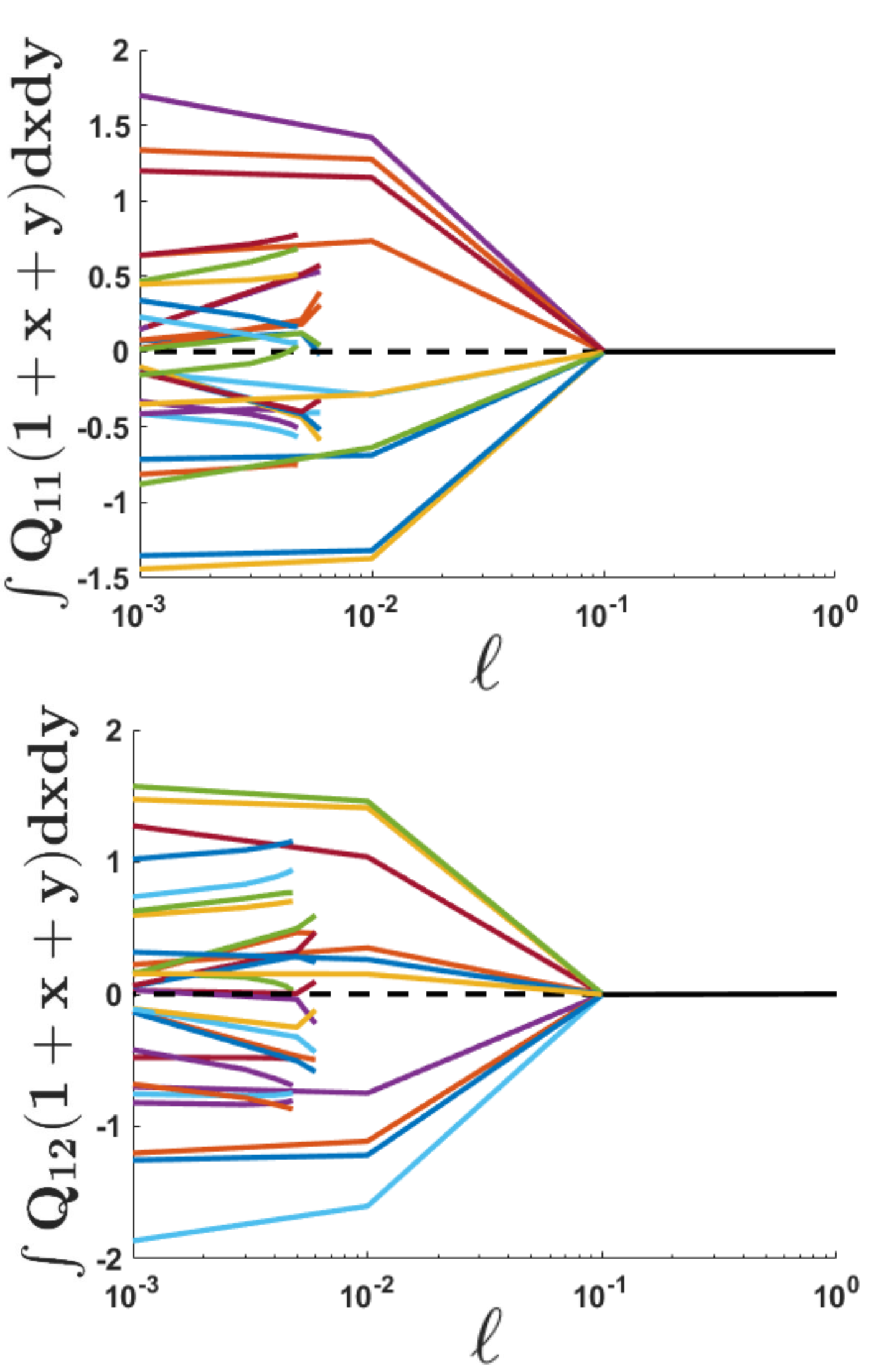}
    \end{subfigure}
        \caption{Bifurcation diagram (\ref{energy}) on a pentagon domain with $c = -0.25$. Left : plot of $\int Q_{11}\left(1+x+y\right)\textrm{d}x\textrm{d}y$, $\int Q_{12}\left(1+x+y\right)\textrm{d}x\textrm{d}y$ versus $\ell$; right: orthogonal 2D projections of the full 3D plot.}
        \label{fig:pentagon_bifurcation_diagram_neg25}
\end{figure}

In Figure~\ref{fig:pentagon_bifurcation_diagram_neg25}, we plot the bifurcation diagram for the $(\Q, \M)$-solutions as a function of $\ell$, for $c= - 0.25$. We lose the symmetry between the splay vertices, and for small $\ell$, we have $5$ stable $Peppa_{in}$, $5$ stable $Peppa_{out}$, $10$ $Meta$ and $10$ $Ortho$ stable solution branches. The symmetry breaking and the preferential perpendicular co-alignment between $\n$ and $\M$ essentially doubles the number of admissible stable states for negative coupling, in the $\ell \to 0$ limit. This provides an ingenious mechanism for stabilising exotic point defects at polygon vertices and in the interior, which could offer novel optical and material responses for future applications.

\section{Conclusions}
In this article, we have studied 2D systems with nematic orientational order and directional magnetic order on regular 2D polygons, with Dirichlet conditions for $\Q$ and $\M$ on the polygon edges. The Dirichlet conditions are special in the sense that we impose a topologically non-trivial boundary condition on $\M$, which necessarily creates an interior magnetic vortex for the $\M$-profiles. Our work is motivated by dilute ferronematic suspensions in 2D frameworks (see \cite{canevari2020design, BishtKonark2019Mnia, bisht20}) and we study observable, physically relevant states in terms of local or global minimizers of an appropriately defined free energy. This approach may apply more widely to model systems with polar and apolar order parameters. The free energy has three contributions - a nematic energy, a magnetic energy and a nemato-magnetic coupling energy. There are four phenomenological parameters in the free energy and with some assumptions, we study the interplay between two parameters: a re-scaled elastic constant $\ell$ and a nemato-magnetic coupling parameter, $c$.
We study the solution landscapes on a 2D square, regular hexagon, regular pentagon in terms of bifurcation diagrams, for $c = 0.25$ and $c = -0.25$ and varying $\ell$. The asymptotics for large $\ell$ are well understood in terms of the $Ring$ branch, since there is a unique critical point/ global minimizer of the free energy in the $\ell \to \infty$ limit. As $\ell$ decreases, the multiplicity of stable $(\Q, \M)$-solutions increases and the solution landscape becomes increasingly complicated. The multistability can be partially understood for small $\ell$ (which correspond to ``large" domains on the micron scale or larger) in terms of a boundary-value problems for $\phi$ and relations between $\theta$ and $\phi$, which define the the nematic director and magnetization vector respectively.

For $c=0$ and $\ell$ small enough, the polygon vertices act as defect sites for stable $\Q$-profiles and in fact, there are at least, $\frac{N(N-1)}{2}$  stable $\Q$-states on a regular polygon of $N$ sides. For positive $c$ that favours co-alignment between $\n$ and $\M$, the number of stable states decreases as $c$ increases, as $\ell \to 0$. In fact, we conjecture that there are only $N$ stable states on a $N$-polygon with $N$ sides for odd $N$, and only $N/2$ stable states for a $N$-polygon with $N$ even, in the $\ell \to 0$ limit, and for large $c$. These stable states are featured by a pair of stable interior $1/2$-nematic point defects in the polygon interior, aligned either parallel to a polygon edge ($N$ odd) or along a polygon diagonal ($N$ even). As $N \to \infty$ and $\ell \to 0$, we recover the solution landscape on a circle with tangent boundary conditions for the $\Q$-profiles: infinitely many stable states with an interior nematic defect pair along one of the circle diagonals \cite{HanYucen2019Tpbd} for $c>0$ and these states cannot be obtained for $c=0$. The $\M$-profiles are less affected in the regime of $c>0$ and small $\ell$, they retain the interior central magnetic vortex with some distortion. We refer to these novel solution branches with interior nematic defect pairs, as $Peppa$ solution branches for $c>0$ and small $\ell$. Informally speaking, positive $c$ has the same effect as regularising the boundary or rounding off the vertices, so that the nematic defects detach from the polygon vertices 
and localise near the polygon centre. Stable domain walls are observed in the $\M$-profile for very small positive values of $c$ or small values of $\xi$, whilst the corresponding $\Q$-profiles retain defects pinned at the polygon vertices. 

The case of $c<0$ that favours $\left(\n \cdot \M \right) = 0$ in the polygon interior, is more complicated. The picture in the $\ell \to \infty$ limit (small nano-scale domains) is qualitatively unchanged in terms of the unique $Ring$ solution branch but $c<0$ strongly enhances multistability in the $\ell \to 0$ limit. We obtain stable solution branches with interior defects for both $\Q$ and $\M$, and additionally, we also find stable solution branches with point defects at the polygon vertices in both the nematic and magnetic profiles. These solution branches with vertex defects, and interior defects, co-exist and could offer exciting optical and electro-magnetic responses to light and external fields. Of course, the experimental tuning of $c$ is expected to be hugely challenging and perhaps a material property, and we expect the case of positive $c$ to be more common in applications than negative $c$. As mentioned in Section~\ref{sec:model}, one might expect $\ell_2 << \ell_1$ for a dilute ferronematic system. We have carried out preliminary numerical investigations by varying the ratio $\frac{\ell_2}{\ell_1}$ with $\ell_1 = 0.005$, $c=0.25$, $\xi =1$. As this ratio decreases from unity, the defects in the $Peppa$-solution branch move towards the vertices and we recover the $Para$-nematic solution branch on a hexagon, which is not attainable for $\ell_2 = \ell_1$ and $c=0.25$. However, we recover the $Peppa$-solution branch for $\ell_2 << \ell_1$ for large enough values of $c$. Hence, we argue that the solution branches for $\ell_1 = \ell_2$ survive for $\ell_2 << \ell_1$, for large values of $c$.

Our study is by no means exhaustive but it does illustrate some generic features of positive and negative $c$, and the roles of $\ell$ and the geometry, in terms of $N$. We do not comment on physical relevance at this stage, but our methods have applications to generic systems with multiple order parameters, of which dilute ferronematics are an example \cite{mertelj2013ferromagnetism, calderer2014effective, ShuaiM2016Slca}. 
Our numerical findings suggest that we will observe mulitstability in this regime, with co-existence of stable solutions supporting a variety of singular structures: magnetic domain walls, stable interior magnetic and nematic defects, boundary vortices, all of which depend on a subtle interplay between $N$, $c$ and $\ell$. It may also be possible to stabilise multiple interior defect pairs, or interior and boundary vortices simultaneously, with a judicious interplay of the model parameters. 
Of course, we have neglected a number of crucial physical considerations e.g., elastic anisotropy, dipolar interactions, weak anchoring, mixed anchoring, the topology of the boundary conditions and flow effects, all of which offer new horizons for complex systems and tailor-made applications.

\section*{Acknowledgments} 
The authors thank the DST-UKIERI for funding the project on ``Theoretical and experimental studies of suspensions of magnetic nanoparticles, their applications and generalizations". The authors also thank Varsha Banerjee and Konark Bisht for working with Apala Majumdar (AM) on ferronematics. AM and YH gratefully ackowledge support from a Royal Society Newton Advanced Fellowship. AM acknowledges support from the University of Strathclyde New Professor's fund and the Leverhulme Trust. YH gratefully acknowledges support from a Royal Society Newton International Fellowship, and thanks Prof. Lei Zhang and Beijing International Center for Mathematical Research of Peking University for hosting her as a Visiting Scholar. AM thanks Samo Kralj for their initial joint work on order reconstruction solutions. 

\bibliographystyle{unsrt}
\bibliography{ferronematics_R1.bib}
	
\end{document}